\documentclass[onecolumn,epjc3]{svjour3}
\usepackage{amsfonts}
\usepackage{mathrsfs}
\usepackage{dblfnote}
\usepackage{graphicx}
\usepackage{hyperref}
\usepackage{bm}
\usepackage{amsmath}
\DeclareMathAlphabet{\mathscrbf}{OMS}{mdugm}{b}{n}
\usepackage{geometry}
\usepackage{indentfirst}
\usepackage{color}
\usepackage{lineno}
\usepackage{cancel}
\usepackage{inputenc}
\usepackage{subfig}
\usepackage{amsmath,graphicx,hyperref,bm,hyperref}
\usepackage{amssymb}
\usepackage[numbers,sort&compress]{natbib}
\usepackage{caption} 
\begin{document}

%\linenumbers

\bibliographystyle{unsrt}

\title{Primordial gravitational waves and curvature perturbations induced energy density perturbation}

\author{
  Zhe Chang\thanksref{addr1,addr2} \and
  Yu-Ting Kuang\thanksref{e1,addr1,addr2} \and
  Xukun Zhang\thanksref{addr1,addr2} \and
  Jing-Zhi Zhou\thanksref{addr1,addr2} 
}
\thankstext{e1}{e-mail: kuangyt@ihep.ac.cn}

\institute{Institute of High Energy Physics, Chinese Academy of Sciences, 100049 Beijing, China\label{addr1}\and 
University of Chinese Academy of Sciences, Beijing 100049, China\label{addr2}
}
\date{Received: date / Accepted: date}

\maketitle

\setlength{\parindent}{1.7em}

\begin{abstract}
    We study the second order scalar and density perturbations generated by the Gaussian curvature perturbations and primordial gravitational waves in the radiation-dominated era. After presenting all the possible second-order source terms, we obtain the explicit expressions of the kernel functions and the power spectra of the second order scalar perturbations. It shows that the primordial gravitational waves might affect the second order energy density perturbation $\delta^{(2)}=\delta \rho^{(2)}/\rho^{(0)}$ significantly. The effects of the primordial gravitational waves are studied in terms of different kinds of primordial power spectra.
\end{abstract}

\section{Introduction}\label{sec:intro}

The cosmological perturbations which are originated from the quantum fluctuations during inflation will inevitably induce higher order perturbations. These induced higher order perturbations can also affect the evolutions of the universe \cite{Planck:2018vyg}.

The cosmological perturbations can be decomposed as scalar, vector, and tensor perturbations on account of the $SO(3)$ symmetry of the Friedmann-Robertson-Walker (FRW) spacetime \cite{Malik:2008im,Mukhanov:1990me,Kodama:1984ziu}. Recently, the higher order perturbations induced by the primordial perturbations have been attracting a lot of interests because of their rich phenomenology \cite{Mollerach:2003nq}. For tensor perturbation, the higher order induced tensor perturbations are known as induced gravitational waves (GWs) \cite{Baumann:2007zm,Kohri:2018awv,Domenech:2021ztg,Chang:2022nzu,Romero-Rodriguez:2021aws,Saito:2008jc,Wang:2019kaf,Inomata:2018epa,Barausse:2020rsu,Bartolo:2018evs,Cai:2018dig,Chang:2022vlv,Inomata:2016rbd,Papanikolaou:2020qtd,Zhou:2020kkf,Domenech:2020kqm,Cai:2019bmk,Cai:2019elf,Yuan:2019udt,Bartolo:2018qqn,Alabidi:2013lya,Zhang:2022dgx,Gong:2019mui,Zhou:2021vcw,LISA:2022kgy,Zhao:2022kvz}. For higher order induced scalar perturbations \cite{Inomata:2020cck,Carrilho:2015cma,Zhang:2017hiu}, the higher order energy density perturbation $\delta^{(n)}=\rho^{(n)}/\rho^{(0)}$ can be calculated in terms of these scalar perturbations. And $\delta^{(n)}$ can affect the primordial black hole (PBH) formation \cite{Nakama:2016enz,Nakama:2015nea}, and the large-scale structure (LSS) of the Universe \cite{Bari:2021xvf,Cho:2020zbh}. The higher order induced vector perturbations can also affect many cosmological observations \cite{Saga:2017hft,Lu:2007cj,Lu:2008ju,Smith:2007sb,Mollerach:2003nq,Acquaviva:2002ud,Kamionkowski:1996zd,Durrer:1994uu,Yamauchi:2012bc,Saga:2015apa,Chang:2022dhh}, such as redshift-space distortions \cite{Smith:2007sb} and weak lensing \cite{Durrer:1994uu,Yamauchi:2012bc,Saga:2015apa}.

The source terms of high order induced perturbations originate from the primordial perturbations generated during the inflation. Since vector perturbations decay as $1/a^2$ \cite{Saga:2017hft}, we typically neglect the primordial vector perturbations. On large scales $(\geq1 Mpc)$, the amplitude of the primordial scalar perturbation $A_{\zeta}$ is constrained by observations of the Cosmic Microwave Background (CMB) and large-scale structures to be about $A_{\zeta}\simeq2\time10^{-9}$. For the primordial tensor perturbation on large scales $(\geq1 Mpc)$, the tensor-to-scalar ratio $r = A_{h}/A_{\zeta}$ is constrained to be less than 0.06 \cite{Planck:2018jri}, where $A_{h}$ is the amplitude of primordial tensor perturbation. Therefore, when studying higher order induced perturbations on large scales $(\geq1 Mpc)$, primordial tensor perturbation can be neglected compared to primordial scalar perturbation.

On small scales $(\leq1 Mpc)$, the constraints of primordial scalar and tensor perturbations are significantly weaker than those on large scales \cite{Bringmann:2011ut}. Over the past few years, the primordial scalar perturbation with large amplitude on small scales have been attracting a lot of interest. It is closely related to primordial black holes and scalar induced GWs \cite{Ananda:2006af,Alabidi:2012ex,Bugaev:2009zh,Inomata:2018cht, Garcia-Bellido:2017aan, Sasaki:2018dmp}. For the primordial tensor perturbation on small scales, its amplitude could also be much larger than it is on the large scales. The large primordial tensor perturbation on small scales can be realized by many models of early universe, such as $G^2$-inflation \cite{Kobayashi:2011nu}, spectator field \cite{Gorji:2023ziy}, and so on \cite{Nakama:2016enz}. Recently, the power spectra of second order tensor perturbation induced by primordial scalar and tensor perturbations with large amplitudes were studied in Ref.~\cite{Chang:2022vlv, Chen:2022dah, Bari:2023rcw, Picard:2023sbz}. They considered the log-normal primordial scalar and tensor power spectra on small scales and found that the primordial tensor perturbation has a very important effect on the second order induced tensor perturbation.

The second order induced scalar and energy density perturbations have been studied for many years \cite{Nakama:2016enz,Nakama:2015nea,Carrilho:2015cma,Zhang:2017hiu,Inomata:2020cck,Bari:2021xvf,Cho:2020zbh,Cho:2022maa}. However, a complete study of the second order induced scalar perturbations has not been presented. The importance of the scalar-tensor coupling source terms $\mathcal{S}_2\sim \phi h$ has been neglected in previous studies. In this paper, we consider the second order energy density perturbation induced by primordial curvature and tensor perturbations during RD era systematically. The second order scalar perturbations can be generated by the scalar-scalar, scalar-tensor, and tensor-tensor coupling source terms: $\mathcal{S}_1\sim \phi\phi$, $\mathcal{S}_2\sim \phi h$, and $\mathcal{S}_3\sim hh$. The second order energy density perturbation $\mathcal{P}^{(2)}_{\delta}$ can be calculated in terms of these second order induced scalar perturbations.The explicit expressions of second order scalar and energy density perturbations are presented in this work.

This paper is organized as follows. In Sec.~\ref{sec:2.0}, the second order scalar perturbations are studied. The explicit expressions of the second order power spectra are presented. In Sec.~\ref{sec:3.0}, we investigate the second order power spectrum in terms of the monochromatic primordial power spectra. The $\mathcal{P}^{(2)}_{\delta}$ induced by  log-normal primordial power spectra are studied in Sec.~\ref{sec:4.0}. The conclusions and discussions are summarized in Sec.~\ref{sec:5}.
%\section{Materials and Methods}

%Materials and Methods should be described with sufficient details to allow others to replicate and build on published results. Please note that publication of your manuscript implicates that you must make all materials, data, computer code, and protocols associated with the publication available to readers. Please disclose at the submission stage any restrictions on the availability of materials or information. New methods and protocols should be described in detail while well-established methods can be briefly described and appropriately cited.

%Research manuscripts reporting large datasets that are deposited in a publicly avail-able database should specify where the data have been deposited and provide the relevant accession numbers. If the accession numbers have not yet been obtained at the time of submission, please state that they will be provided during review. They must be provided prior to publication.

%Interventionary studies involving animals or humans, and other studies require ethical approval must list the authority that provided approval and the corresponding ethical approval code.

\section{Second order scalar perturbations}\label{sec:2.0}
In this section, we study the equations of motion and the kernel functions of the second order scalar perturbations induced by primordial curvature and tensor perturbations in the RD era.

\subsection{Equation of motion}\label{sec:2}
\begin{figure}[htbp]
	\centering
	\subfloat[$\mathcal{S}_{ij,1}\sim\phi^{(1)}\phi^{(1)}$]{\includegraphics[scale = 0.8]{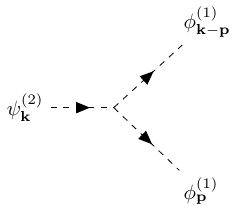}}
	\subfloat[$\mathcal{S}_{ij,2}\sim\phi^{(1)} h^{\lambda,(1)}$]{\includegraphics[scale = 0.8]{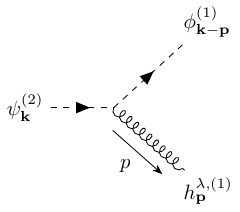}}
	\subfloat[$\mathcal{S}_{ij,3}\sim h^{\lambda,(1)} h^{\lambda,(1)}$]{\includegraphics[scale = 0.8]{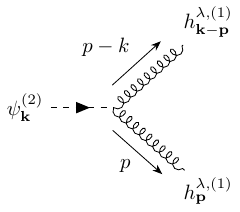}}
	\caption{ The source terms $\mathcal{S}_{ij,1}$ is composed of the first order scalar perturbation, the source terms $\mathcal{S}_{ij,2}$ is composed of the product of the first order scalar perturbation and the first order tensor perturbation, and the source terms $\mathcal{S}_{ij,3}$ is composed of the first order tensor perturbation.}\label{fig:fey}
\end{figure}

The perturbed metric in the flat FRW spacetime with Newtonian gauge is given by
\begin{equation}
	\begin{aligned}
		\mathrm{d} s^{2}&=a^{2}\Bigg[-\left(1+2 \phi^{(1)}+\phi^{(2)}\right) \mathrm{d} \eta^{2}+\left(\left(1-2 \psi^{(1)}-\psi^{(2)}\right) \delta_{i j}+h^{(1)}_{ij}\right)\mathrm{d} x^{i} \mathrm{d} x^{j}\Bigg] \ ,
	\end{aligned}
\end{equation}
where $\phi^{(n)}$ and $\psi^{(n)}$$\left( n=1,2 \right)$ are first order and second order scalar perturbations, $h^{(1)}_{ij}$ is the first order tensor perturbation. Here, we neglect the first order vector perturbation since the vector modes decay as $1/a^2$ after
they leaving the Hubble horizon during inflation. We use \texttt{xPand} package to study the perturbations of Einstein equation, \texttt{xPand} package can help us to obtain and simplify the equations of motion of cosmological perturbations \cite{Pitrou:2013hga}. The equations of motion of second order scalar perturbations are given by
\begin{eqnarray}\label{eq:2.2}
	\Psi^{(2)}-\Phi^{(2)}=-2 \Delta^{-1}\left(\partial^i \Delta^{-1} \partial^j-\frac{1}{2} \mathcal{T}^{i j}\right) \mathcal{S}_{i j}\left(\mathbf{x},\eta  \right) \ ,
	\label{eq:eq1}
\end{eqnarray}
\begin{eqnarray}
		\partial_{\eta}^2 \Psi^{(2)}+3\mathcal{H} \partial_{\eta} \Psi^{(2)}-\frac{5}{6} \Delta \Psi^{(2)}+\mathcal{H} \partial_{\eta} \Phi^{(2)}+\frac{1}{2} \Delta \Phi^{(2)}=-\frac{1}{2} \mathcal{T}^{i j} \mathcal{S}_{i j}\left(\mathbf{x},\eta  \right) \ ,
			\label{eq:eq2}
\end{eqnarray}
where $\mathcal{T}^{i j}=\delta^{ij}-\partial^i\Delta^{-1} \partial^j$ is the transverse operator. During the RD era, the conformal Hubble parameter can be expressed as $\mathcal{H}=1/\eta$. For convenience, we use the symbols $\phi^{(1)}\equiv \phi$ and $h_{ij}^{(1)}\equiv h_{ij}$. As shown in Fig.~\ref{fig:fey}, the source term $\mathcal{S}_{ij}\left(\mathbf{x},\eta  \right)$ is composed of three parts $\mathcal{S}_{ij}\left(\mathbf{x},\eta  \right) =\mathcal{S}_{ij,1}+\mathcal{S}_{ij,2}+\mathcal{S}_{ij,3} $.
The explicit expressions of the source terms are given in~\ref{sec:S}. Substituting Eq.~(\ref{eq:eq1}) into Eq.~(\ref{eq:eq2}), we obtain the equation of motion of second order scalar perturbation $\Psi^{(2)}$ in the RD era 
\begin{equation}\label{eq:eq3}
	\begin{aligned}
		&\partial^2_{\eta} \Psi^{(2)}+\frac{4}{\eta} \partial_{\eta} \Psi^{(2)}-\frac{1}{3}\Delta \Psi^{(2)} \\
		&=-\frac{1}{2} \mathcal{T}^{ij} S_{ij}-2 \Delta^{-1}\left(\partial^j \Delta^{-1} \partial^i-\frac{1}{2} \mathcal{T}^{ij}\right)\left(\frac{1}{2} \Delta+\frac{1}{\eta} \partial_{\eta}\right) \mathcal{S}_{ij}\left(\mathbf{x},\eta  \right)\\
		&=\left(-\frac{1}{2} \mathcal{T}^{ij} -\left(\partial^j \Delta^{-1} \partial^i-\frac{1}{2} \mathcal{T}^{ij}\right)-2 \Delta^{-1}\left(\partial^j \Delta^{-1} \partial^i-\frac{1}{2} \mathcal{T}^{ij}\right)\frac{1}{\eta} \partial_{\eta}\right) S_{ij}\left(\mathbf{x},\eta  \right) \\
		&=-\left(\partial^j \Delta^{-1} \partial^i+2 \Delta^{-1}\left(\partial^j \Delta^{-1} \partial^i-\frac{1}{2} \mathcal{T}^{ij}\right)\frac{1}{\eta} \partial_{\eta}\right) \sum_{a=1}^{3}\mathcal{S}_{ij,a}\left(\mathbf{x},\eta  \right) \ .
	\end{aligned} 
\end{equation}
\subsection{Kernel functions}\label{sec:3}
\begin{figure}
	\includegraphics[scale = 0.72]{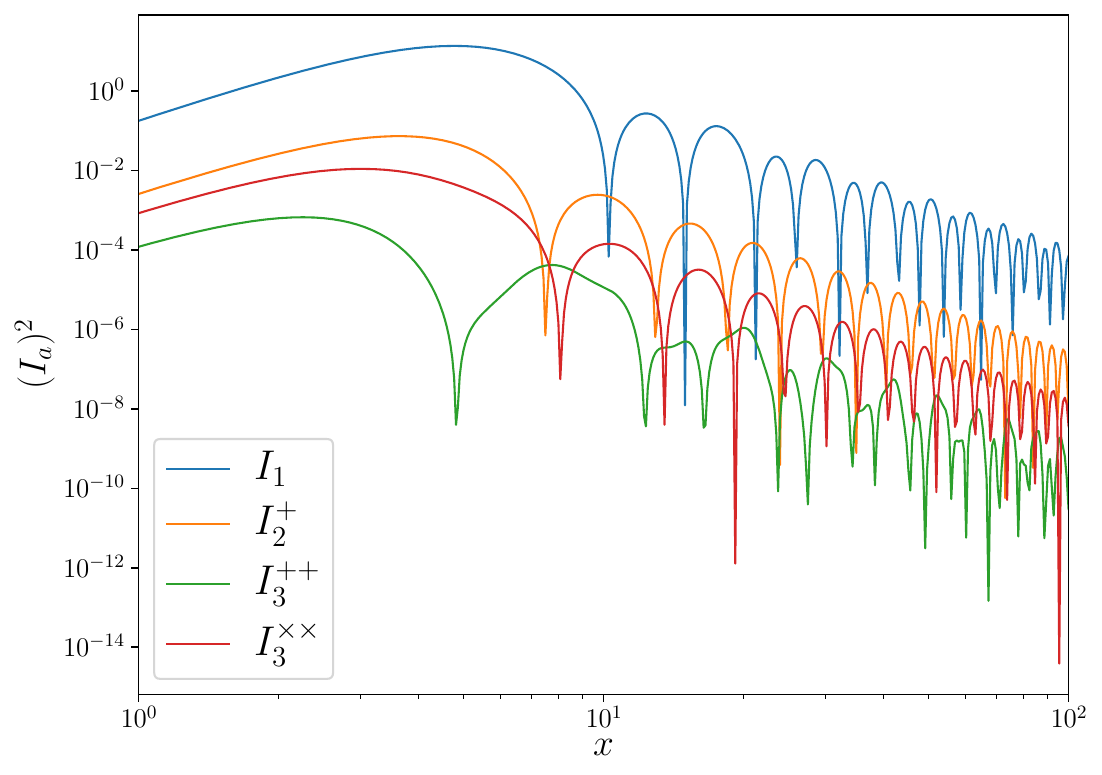}
	\caption{The kernel functions $I_i\ (i=1,2,3)$ in Eq.~(\ref{eq:I}). We have set $u=v=1$. }\label{fig:I}
\end{figure}

In order to solve the equation of motion of second order scalar perturbation, we rewrite Eq.~(\ref{eq:eq3}) in momentum space as
\begin{equation}\label{eq:eqk}
	\begin{aligned}
		 \Psi^{(2)''}\left(\mathbf{k},\eta  \right)+\frac{4}{\eta}\Psi^{(2)'}\left(\mathbf{k},\eta  \right)+\frac{k^2}{3} \Psi^{(2)}\left(\mathbf{k},\eta  \right)=\sum_{a=1}^{3}\mathcal{S}_{a}\left(\mathbf{k},\eta  \right)  \ ,
	\end{aligned} 
\end{equation}
where 
\begin{equation}\label{eq:D}
	\begin{aligned}
		 \mathcal{S}_{a}\left(\mathbf{k},\eta  \right) =\mathcal{D}^{ij}_{\psi}\mathcal{S}_{ij,a}\left(\mathbf{k},\eta  \right) \ , \ \ \mathcal{D}_{\psi}^{ij}=-\left(\frac{k^ik^j}{k^2}-\left(\frac{3k^ik^j}{k^4}-\frac{\delta^{ij}}{k^2}  \right)\frac{k^2}{x}\partial_x  \right)\ .
	\end{aligned} 
\end{equation}
Here, we have defined $x\equiv k\eta$. The explicit expressions of  $\mathcal{S}_{ij,a}\left(\mathbf{k},\eta  \right)$$(a=1,2,3)$ are given in~\ref{sec:S}. 
Substituting Eqs.~(\ref{eq:s1})--(\ref{eq:s3}) into Eq.~(\ref{eq:D}), we obtain the expressions of $\mathcal{S}_{a}\left(\mathbf{k},\eta  \right)$
\begin{eqnarray}
	\mathcal{S}_1&=&\mathcal{D}^{ij}_{\psi}\mathcal{S}_{ij,1}=\int\frac{d^3p}{(2\pi)^{3/2}}k^2f_1\left(u,v,x \right) \frac{4}{9}\zeta_{\mathbf{k}-\mathbf{p}}\zeta_{\mathbf{p}} \ ,
	\label{eq:f1} \\
	\mathcal{S}_2&=&\mathcal{D}^{ij}_{\psi}\mathcal{S}_{ij,2}=\int\frac{d^3p}{(2\pi)^{3/2}}k^2f^{\lambda_1}_2\left(u,v,x \right)\frac{2}{3} \zeta_{\mathbf{k}-\mathbf{p}}\mathbf{h}^{\lambda_1}_{\mathbf{p}} \ ,
	\label{eq:f2} \\
	\mathcal{S}_3&=&\mathcal{D}^{ij}_{\psi}\mathcal{S}_{ij,3}=\int\frac{d^3p}{(2\pi)^{3/2}}k^2f^{\lambda_1\lambda_2}_3\left(u,v,x \right) \mathbf{h}^{\lambda_1}_{\mathbf{k}-\mathbf{p}}\mathbf{h}^{\lambda_2}_{\mathbf{p}} \ ,
	\label{eq:f3} 
\end{eqnarray}
where $\lambda_1$ and $\lambda_2$ are the polarization indices, the spatial indices of $\mathcal{S}_{ij,a}$$(a=1,2,3)$ are contracted. Substituting Eqs.~(\ref{eq:f1})--(\ref{eq:f3}) into Eq.~(\ref{eq:eqk}), we solve the Eq.~(\ref{eq:eqk}) by using the Green's function method
\begin{equation}\label{eq:Psi}
\Psi^{(2)}=\sum_{a=1}^{3} \Psi^{(2)}_a  \ , \ (a=,1,2,3) \ ,
\end{equation}
where
\begin{eqnarray}
	\Psi^{(2)}_1&=&\int\frac{d^3p}{(2\pi)^{3/2}}I_1\left(u,v,x \right) \frac{4}{9}\zeta_{\mathbf{k}-\mathbf{p}}\zeta_{\mathbf{p}} \ ,
	\label{eq:psi1} \\
	\Psi^{(2)}_2&=&\int\frac{d^3p}{(2\pi)^{3/2}}I^{\lambda_1}_2\left(u,v,x \right)\frac{2}{3} \zeta_{\mathbf{k}-\mathbf{p}}\mathbf{h}^{\lambda_1}_{\mathbf{p}} \ ,
	\label{eq:psi2} \\
	\Psi^{(2)}_3&=&\int\frac{d^3p}{(2\pi)^{3/2}}I^{\lambda_1\lambda_2}_3\left(u,v,x \right) \mathbf{h}^{\lambda_1}_{\mathbf{k}-\mathbf{p}}\mathbf{h}^{\lambda_2}_{\mathbf{p}} \ .
	\label{eq:psi3} 
\end{eqnarray}
In Eqs.~(\ref{eq:psi1})--(\ref{eq:psi3}), the kernel functions $I_{a}(u, v, x)$$(a=1,2,3)$ are defined as
\begin{equation} \label{eq:I}
	\begin{aligned}
		I_{1}&=\int_0^x \mathrm{~d} \bar{x}\left(\frac{\bar{x}}{x}\right)^2\left\{-\frac{x \bar{x}}{\sqrt{3}}\left[j_1(x / \sqrt{3}) y_1(\bar{x} / \sqrt{3})-j_1(\bar{x} / \sqrt{3}) y_1(x / \sqrt{3})\right]\right\} f_1(u, v, \bar{x}) \ , \\
		I_{2}^{\lambda_1}&=\int_0^x \mathrm{~d} \bar{x}\left(\frac{\bar{x}}{x}\right)^2\left\{-\frac{x \bar{x}}{\sqrt{3}}\left[j_1(x / \sqrt{3}) y_1(\bar{x} / \sqrt{3})-j_1(\bar{x} / \sqrt{3}) y_1(x / \sqrt{3})\right]\right\} f^{\lambda_1}_2(u, v, \bar{x}) \ , \\
		I_{3}^{\lambda_1,\lambda_2}&=\int_0^x \mathrm{~d} \bar{x}\left(\frac{\bar{x}}{x}\right)^2\left\{-\frac{x \bar{x}}{\sqrt{3}}\left[j_1(x / \sqrt{3}) y_1(\bar{x} / \sqrt{3})-j_1(\bar{x} / \sqrt{3}) y_1(x / \sqrt{3})\right]\right\} f^{\lambda_1,\lambda_2}_3(u, v, \bar{x}) \ .
	\end{aligned}
\end{equation} 
In the end of this section, we calculate the kernel functions in Eq.~(\ref{eq:I}). We present these three kinds of kernel functions  $\left(I_a(u=1,v=1,x)\right)^2$$(a=1,2,3)$ as functions of $x=k\eta$ in Fig.~\ref{fig:I}.
As shown in Fig.~\ref{fig:I}, the kernel function $I_1$ is much larger than other kernel functions.
\subsection{Initial second order perturbation}
As we mentioned in Ref.~\cite{Malik:2008im,Inomata:2020cck}, the contributions from the initial second-order perturbation also need to considered. More precisely, the second order scalar perturbations are composed of two parts, the second order scalar perturbations induced by the primordial perturbations and the initial second-order perturbation. The second-order curvature perturbation in the Newtonian gauge is given by \cite{Malik:2008im}
\begin{equation}
	\begin{aligned}
		-\zeta^{(2)}=\Psi^{(2)}+\frac{\mathcal{H}}{\rho^{(0) \prime}}\left[\delta \rho^{(2)}-\frac{\delta \rho^{(1) \prime}}{\rho^{(0) \prime}} \delta \rho^{(1)}\right]-\frac{1}{4} \chi_{\delta \rho k}^k+\frac{1}{4} \Delta^{-2}\partial^i\partial^j \chi_{ij \delta \rho } \ ,
	\end{aligned}
\end{equation}
where
\begin{equation}
	\begin{aligned}
		\chi_{i j \delta \rho}& \equiv  -2 \frac{\mathcal{H}}{\rho^{(0) \prime}}\left[2 \mathcal{H}\left(\frac{\delta \rho^{(1) 2}}{\rho^{(0) \prime}}\right)-\frac{\delta \rho^{(1) \prime}}{\rho^{(0) \prime}} \delta \rho^{(1)}\right] \delta_{i j}-\frac{2}{\rho^{(0) ' 2}} \partial_i\delta \rho^{(1)}\partial_j \delta \rho^{(1)} \\
		& +4\left[-\frac{\delta \rho^{(1)}}{\rho^{(0) \prime}}\left[\left(-\Psi^{(1) \prime} \delta_{i j}+\frac{1}{2} h_{i j}^{(1) \prime}\right)+2 \mathcal{H}\left(-\Psi^{(1)} \delta_{i j}+\frac{1}{2} h_{i j}^{(1)}\right)\right]\right]   \ , 
	\end{aligned}
\end{equation}
\begin{equation}
	\begin{aligned}
		\chi_{ \delta  \rho k}^k &\equiv  -6 \frac{\mathcal{H}}{\rho^{(0) \prime}}\left[2 \mathcal{H}\left(\frac{\delta \rho^{(1) 2}}{\rho^{(0) \prime}}\right)-\frac{\delta \rho^{(1) \prime}}{\rho^{(0) \prime}} \delta \rho^{(1)}\right]-\frac{2}{\rho^{(0) ' 2}}\partial^k \delta \rho^{(1)}\partial_k \delta \rho^{(1)} \\
		& +4\left[-\frac{\delta \rho^{(1)}}{\rho^{(0) \prime}}\left[\left(-3 \Psi^{(1) \prime}+\frac{1}{2} h_k^{(1) k \prime}\right)+2 \mathcal{H}\left(-3 \Psi^{(1)}+\frac{1}{2} h_k^{(1) k}\right)\right]\right]  \ .
	\end{aligned}
\end{equation}
In the  superhorizon limit ($k\eta \ll 1$), the second-order curvature perturbation can be approximated as
\begin{equation}\label{eq:sl}
	\begin{aligned}
		\zeta^{(2)}(k, \eta) \simeq & \left\{-\Psi^{(2)}(k, \eta)+\frac{1}{4} \delta^{(2)}-\frac{1}{8} \delta^{(1) 2}-\delta^{(1)} \Psi^{(1)}\right. \\
		& \left.+\frac{\delta^{(1)} h_k^{(1) k}}{4}-\frac{\delta^{(1)} \nabla^{-2} \partial_i \partial_j h^{(1) i j}}{4}\right\} T_{\Psi}(k \eta) \ ,
	\end{aligned}
\end{equation}
where
\begin{equation}
	\begin{aligned}
		\delta^{(1)}=\frac{\delta \rho^{(1)}}{\rho^{(0)}}=\frac{-6 \mathcal{H}\left(\mathcal{H} \phi^{(1)}+\psi^{(1) \prime}\right)+2 \triangle \psi^{(1)}}{3 \mathcal{H}^2} \simeq-2 \phi^{(1)}=-\frac{4}{3} \zeta^{(1)} \ .
	\end{aligned}
\end{equation}
We assume the local type non-Gaussianity here, which is parametrized as $\zeta^{(2)}=2a_{\mathrm{NL}}\left(\zeta^{(1)}\right)^2$ in the superhorizon limit, and $a_{\mathrm{NL}}=1$ for Gaussian perturbation \cite{Bartolo:2004if,Bartolo:2004ty,Young:2015kda}. Substituting the condition of the local type non-Gaussianity into Eq.~(\ref{eq:sl}), we obtain the contributions from the initial second-order perturbation
\begin{equation}\label{eq:pin}
	\begin{aligned}
		\Psi_{\mathrm{ in}}^{(2)}\left( \mathbf{k} \right)&=\frac{1}{3}\left( \frac{3k^ik^j}{k^4}-\frac{\delta^{ij}}{k^2} \right)\mathcal{S}_{ij}+ \int \frac{d^3p}{(2\pi)^{3/2}} \left(\left(-\frac{4}{3}a_{\mathrm{NL}}+\frac{28}{27} \right)\zeta_{\mathbf{k}-\mathbf{p}}\zeta_{\mathbf{p}} \right. \\
		&\left. -\frac{2}{9}\varepsilon_{i}^{\lambda,i}\left(\mathbf{p}\right)\zeta_{\mathbf{k}-\mathbf{p}}\mathbf{h}^{\lambda}_{\mathbf{p}}+\frac{2}{9}\frac{p^ip^j}{p^2}\varepsilon_{ij}^{\lambda}\left(\mathbf{p}\right)\zeta_{\mathbf{k}-\mathbf{p}}\mathbf{h}^{\lambda}_{\mathbf{p}} \right) \ .
	\end{aligned}
\end{equation}
After considering the effects of the initial second-order perturbation, the second order scalar perturbation $\Psi^{(2)}$ can be written as
\begin{equation}\label{eq:ptot}
	\begin{aligned}
		\Psi^{(2)}\left( \mathbf{k} \right)=\Psi_{\mathrm{ in}}^{(2)}\left( \mathbf{k} \right)+\sum_{a=1}^{3} \Psi^{(2)}_a  \ , \ (a=,1,2,3)  \ .
	\end{aligned}
\end{equation}
We study the second order scalar and density perturbations generated by the Gaussian curvature and tensor perturbations. Therefore, we set $a_{\mathrm{NL}}=1$ in this paper.

\subsection{Power spectra}\label{sec:4}
\begin{figure}[htbp]
 	\centering
 	\subfloat[$	\mathcal{P}^{(2)}_{1}\sim	\langle	\Psi_{1}^{(2)}\left( \mathbf{k} \right)\Psi_{1}^{(2)}\left( \mathbf{k}' \right) \rangle$]{\includegraphics[scale = 0.8]{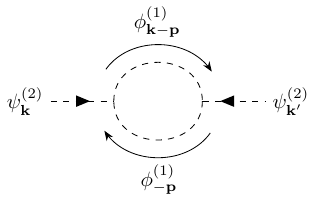}}
 	\subfloat[$\mathcal{P}^{(2)}_{2}\sim	\langle	\Psi_{2}^{(2)}\left( \mathbf{k} \right)\Psi_{2}^{(2)}\left( \mathbf{k}' \right) \rangle$]{\includegraphics[scale = 0.8]{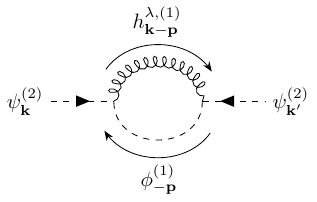}}
 	\subfloat[$\mathcal{P}^{(2)}_{3}\sim	\langle	\Psi_{3}^{(2)}\left( \mathbf{k} \right)\Psi_{3}^{(2)}\left( \mathbf{k}' \right) \rangle$]{\includegraphics[scale = 0.8]{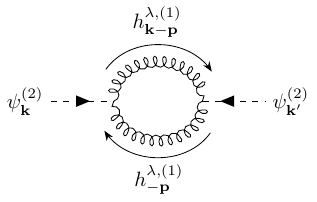}}
 	\caption{The Feynman diagrams of three kinds of two-point functions.}\label{fig:ffey}
 \end{figure}

In this section, the power spectra of second order scalar and density perturbations are investigated.  We assume that the two-point function $\langle \zeta_{\mathbf{k}_1}\mathbf{h}^{\lambda}_{\mathbf{k}_2} \rangle$=0 for arbitrary $\mathbf{k}_1$ and $\mathbf{k}_2$ \cite{Chang:2022vlv}. Therefore, we only need to consider three kinds of four-point functions. The explicit expressions of these four-point functions are given in~\ref{sec:B}. The power spectra of second order scalar perturbation is defined as
\begin{equation}\label{eq:P}
	\begin{aligned}
		\langle	\Psi^{(2)}\left( \mathbf{k} \right)\Psi^{(2)}\left( \mathbf{k}' \right) \rangle=\delta\left(\mathbf{k}+\mathbf{k}' \right)\frac{2\pi^2}{k^3}\mathcal{P}^{(2)}_{\psi} \ .
	\end{aligned} 
\end{equation}
Substituting Eqs.~(\ref{eq:psi1})--(\ref{eq:psi3}) into Eq.~(\ref{eq:P}), we obtain 
\begin{equation}\label{eq:P0}
	\begin{aligned}
		\langle	\Psi^{(2)}\left( \mathbf{k} \right)\Psi^{(2)}\left( \mathbf{k}' \right) \rangle=\sum_{i=1}^{3}\langle	\Psi_{i}^{(2)}\left( \mathbf{k} \right)\Psi_{i}^{(2)}\left( \mathbf{k}' \right) \rangle \ ,
	\end{aligned} 
\end{equation}
where
	\begin{eqnarray}
		\langle	\Psi_{1}^{(2)}\left( \mathbf{k} \right)\Psi_{1}^{(2)}\left( \mathbf{k}' \right) \rangle&=&\int\frac{d^3p}{(2\pi)^{3/2}}\frac{d^3p'}{(2\pi)^{3/2}} I_{1}\left(|\mathbf{k}-\mathbf{p}|,p,\eta  \right)I_{1}\left(|\mathbf{k}'-\mathbf{p}'|,p',\eta  \right) \nonumber\\
		& &\times~\frac{16}{81} \langle \zeta_{\mathbf{k}-\mathbf{p}}\zeta_{\mathbf{p}}\zeta_{\mathbf{k}'-\mathbf{p}'}\zeta_{\mathbf{p}'} \rangle \ ,
		\label{eq:P1} \\
		\langle	\Psi_{2}^{(2)}\left( \mathbf{k} \right)\Psi_{2}^{(2)}\left( \mathbf{k}' \right) \rangle&=&\int\frac{d^3p}{(2\pi)^{3/2}}\frac{d^3p'}{(2\pi)^{3/2}} I^{\lambda_1}_{2}\left(|\mathbf{k}-\mathbf{p}|,p,\eta  \right)I^{\lambda'_1}_{2}\left(|\mathbf{k}'-\mathbf{p}'|,p',\eta  \right) \nonumber\\
		& &\times~\frac{4}{9} \langle \zeta_{\mathbf{k}-\mathbf{p}}\mathbf{h}^{\lambda_1}_{\mathbf{p}}\zeta_{\mathbf{k}'-\mathbf{p}'}\mathbf{h}^{\lambda'_1}_{\mathbf{p}'} \rangle \ ,
		\label{eq:P2} \\
		\langle	\Psi_{3}^{(2)}\left( \mathbf{k} \right)\Psi_{3}^{(2)}\left( \mathbf{k}' \right) \rangle&=&\int\frac{d^3p}{(2\pi)^{3/2}}\frac{d^3p'}{(2\pi)^{3/2}} I^{\lambda_1\lambda_2}_{3}\left(|\mathbf{k}-\mathbf{p}|,p,\eta  \right)I^{\lambda'_1\lambda'_2}_{3}\left(|\mathbf{k}'-\mathbf{p}'|,p',\eta  \right) \nonumber\\
		& &\times~ \langle \mathbf{h}^{\lambda_1}_{\mathbf{k}-\mathbf{p}}\mathbf{h}^{\lambda_2}_{\mathbf{p}}\mathbf{h}^{\lambda'_1}_{\mathbf{k}'-\mathbf{p}'}\mathbf{h}^{\lambda'_2}_{\mathbf{p}'} \rangle \ .
		\label{eq:P3} 
	\end{eqnarray} 
The corresponding Feynman diagrams of these three kinds of two-point functions are given in Fig.~\ref{fig:ffey}
Substituting Eqs.~(\ref{eq:B1})--(\ref{eq:B3}) into Eqs.~(\ref{eq:P1})--(\ref{eq:P3}), we obtain the explicit expressions of the power spectra
\begin{equation}\label{eq:p0}
	\begin{aligned}
		\mathcal{P}^{(2)}_{\psi}=\mathcal{P}^{(2)}_{1}+\mathcal{P}^{(2)}_{2}+\mathcal{P}^{(2)}_{3} \ ,
	\end{aligned} 
\end{equation}
where
	\begin{eqnarray}
		\mathcal{P}^{(2)}_{1}&=&\frac{1}{2}\int^{\infty}_0 dv \int^{|1+v|}_{|1-v|} \frac{du}{v^2u^2}I_{1}\left(u,v,x \right) \mathcal{P}_{\zeta}(uk)\mathcal{P}_{\zeta}(vk) \nonumber\\
		&&\times~\frac{16}{81}\left[ I_{1}\left(|\mathbf{k}'-\mathbf{p}'|,p',\eta \right)\big|_{ \mathbf{p}'=-\mathbf{p}}+I_{1}\left(|\mathbf{k}'-\mathbf{p}'|,p',\eta \right)\big|_{\mathbf{p}'=\mathbf{p}-\mathbf{k}} \right]\Big|_{\mathbf{k}'\to -\mathbf{k}}  \ , 
		\label{eq:p1}\\
		\mathcal{P}^{(2)}_{2}&=&\frac{1}{2}\int^{\infty}_0 dv \int^{|1+v|}_{|1-v|} \frac{du}{v^2u^2}I^{\lambda_1}_{2}\left(u,v,x \right)\delta^{\lambda_1\lambda'_1} \mathcal{P}_{\zeta}(uk)\mathcal{P}_{h}(vk) \nonumber\\
		&&~~~~\times\frac{4}{9}I^{\lambda'_1}_{2}\left(|\mathbf{k}'-\mathbf{p}'|,p',\eta \right)\big|_{ \mathbf{p}'=-\mathbf{p},\ \mathbf{k}'=-\mathbf{k}}\ , 
		\label{eq:p2}\\
		\mathcal{P}^{(2)}_{3}&=&\frac{1}{2}\int^{\infty}_0 dv \int^{|1+v|}_{|1-v|} \frac{du}{v^2u^2}I^{\lambda_1\lambda_2}_{3}\left(u,v,x \right)\mathcal{P}_{h}(uk)\mathcal{P}_{h}(vk) \nonumber\\
		&&~~~~~~\times\left[\delta^{\lambda_1\lambda'_1}\delta^{\lambda_2\lambda'_1}I^{\lambda'_1\lambda'_2}_{3}\left(|\mathbf{k}'-\mathbf{p}'|,p',\eta  \right)\big|_{\mathbf{p}'=-\mathbf{p}} \right. \nonumber\\
		& & ~~~~~~~~~~~~~~~~~~~~\left. +\delta^{\lambda_1\lambda'_2}\delta^{\lambda_2\lambda'_1}I^{\lambda'_1\lambda'_2}_{3}\left(|\mathbf{k}'-\mathbf{p}'|,p',\eta \right)\big|_{\mathbf{p}'=\mathbf{p}-\mathbf{k}}  \right]\Big|_{\mathbf{k}'=-\mathbf{k}} \ .
		\label{eq:p3}
	\end{eqnarray} 
In Eqs.~(\ref{eq:p1})--(\ref{eq:p3}), the substitutions of $\mathbf{k}'$ and $\mathbf{p}'$ come from the three dimensional delta functions in the Wick's expansions of four-point functions.  Since we have assumed the two-point function $\langle \zeta_{\mathbf{k}_1}\mathbf{h}^{\lambda}_{\mathbf{k}_2} \rangle$=0, three kinds of source terms in Eqs.~(\ref{eq:S1})--(\ref{eq:S3}) are decoupled. More precisely, the two-point functions $\langle	\Psi_{i}^{(2)}\left( \mathbf{k} \right)\Psi_{j}^{(2)}\left( \mathbf{k}' \right) \rangle=0$ if $i\ne j$. As shown in Eqs.~(\ref{eq:p1})--(\ref{eq:p3}), the power spectra of second order scalar perturbation are composed of three parts, which come from the source terms $\mathcal{S}_1\sim \phi\phi$, $\mathcal{S}_2\sim \phi h$, and $\mathcal{S}_3\sim hh$ respectively. 

The energy density perturbation can be calculated as
\begin{equation}\label{eq:rho}
	\begin{aligned}
		\delta^{(2)}&=\frac{\delta \rho^{(2)}}{\rho^{(0)}}=-\frac{1}{3\mathcal{H}^4}\Bigg(\frac{1}{4}\mathcal{H}^2h'_{ij}\left( h^{'ij}+8\mathcal{H}h^{ij} \right)+6\mathcal{H}^4\left( -4\phi^2+\Phi^{(2)} \right) +6\mathcal{H}^3\Psi^{(2)'}+4\mathcal{H}\partial_i\phi'\partial^i \phi \\
		&+2\partial_i\phi'\partial^i\phi'+\mathcal{H}^2\left(-2\left(3\phi'^2+8\phi\Delta \phi +\Delta \Psi^{(2)}+2\partial_i\phi\partial^i \phi +\frac{1}{2}h^{ij}\left( -\partial_i\partial_j\phi +\Delta h_{ij} \right)  \right) \right. \\
		&\left. +\frac{1}{4}\left( 2\partial_lh_{ij}-3\partial_j h_{il} \right)\partial^j h^{il}  \right) \Bigg) \ .
	\end{aligned}
\end{equation}
The power spectrum of second order energy density perturbation $\mathcal{P}^{(2)}_{\delta}$ is defined by
\begin{equation}\label{eq:delta}
	\begin{aligned}
		\langle	\delta^{(2)}\left( \mathbf{k} \right)\delta^{(2)}\left( \mathbf{k}' \right) \rangle=\delta\left(\mathbf{k}+\mathbf{k}' \right)\frac{2\pi^2}{k^3}\mathcal{P}^{(2)}_{\delta} \ ,
	\end{aligned} 
\end{equation}
where the energy density $\delta^{(2)}$ can be calculated in terms of Eq.~(\ref{eq:ptot}) and Eq.~(\ref{eq:rho}).

\section{Monochromatic primordial power spectra}\label{sec:3.0}
As mentioned, the constraints of primordial curvature and tensor perturbations on small scales are significantly weaker than those on large scales, the tensor-to-scalar ratio r can be very large on small scales. Therefore, we start with a toy model of $\delta$-peak. In this case, the primordial scalar and tensor perturbations are very large on small scalar. Since we consider the second order scalar and density perturbations generated by the Gaussian curvature and tensor perturbations, we have set $a_{\mathrm{NL}}=1$ in the Eq.~(\ref{eq:pin}).

\subsection{Monochromatic primordial power spectra with the same $k_*$}
\begin{figure}
	\includegraphics[scale = 0.33]{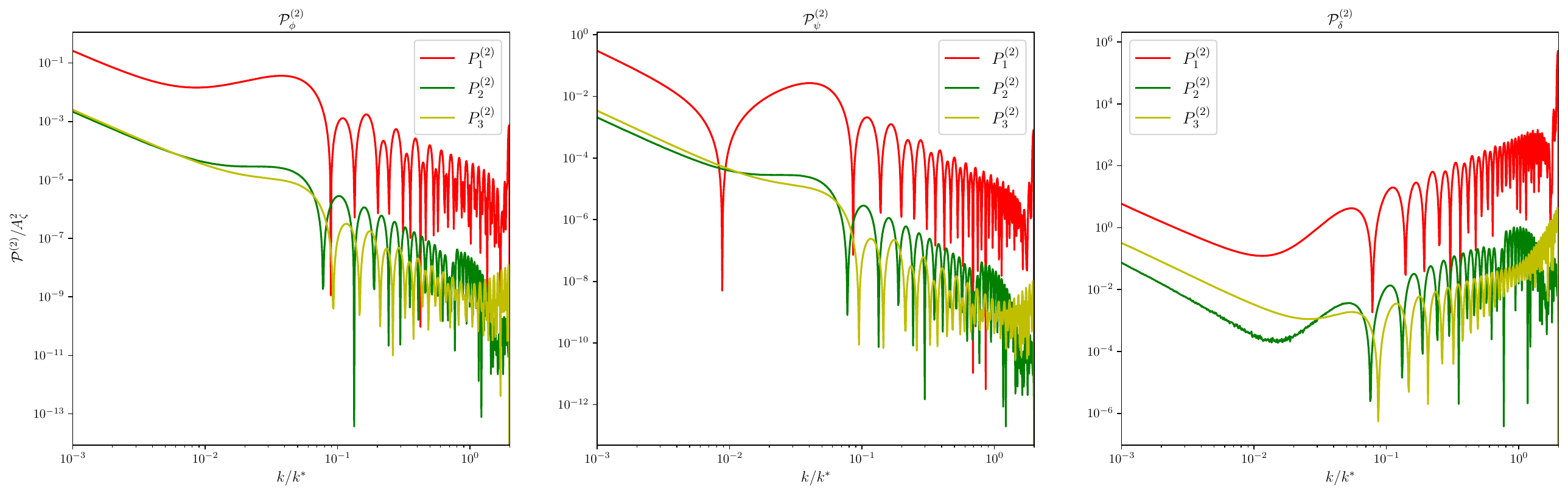}
	\caption{The three kinds of power spectra $\mathcal{P}_a^{(2)}$$(a=1,2,3)$ for second order perturbations $\Phi^{(2)}$, $\Psi^{(2)}$, and $\delta^{(2)}$. The power spectra  $\mathcal{P}_1^{(2)}$, $\mathcal{P}_2^{(2)}$, and $\mathcal{P}_3^{(2)}$ come from the source terms  $\mathcal{S}_1\sim \phi\phi$, $\mathcal{S}_2\sim \phi h$, and $\mathcal{S}_3\sim hh$ respectively.  We have set the tensor-to-scalar ratio $r=A_{h}/A_{\zeta}=1$, and $x_*=k_*\eta=100$. }\label{fig:P_1}
\end{figure}

 We consider the monochromatic primordial power spectra, namely
\begin{equation}
	\begin{aligned}
		\mathcal{P}_{\zeta}=A_{\zeta}k_{\zeta *}\delta\left( k-k_{\zeta *} \right) \ , \ \mathcal{P}_{h}=A_{h}k_{h*}\delta\left( k-k_{h*} \right) \ , \  k_{\zeta *}=k_{h *}=k_* \  ,
	\end{aligned}
\end{equation}
where $k_{*}$ is the wavenumber at which the power spectrum has a $\delta$-function peak.  In Fig.~\ref{fig:P_1}, we plot the three kinds of power spectra for second order perturbations $\Phi^{(2)}$, $\Psi^{(2)}$, and $\delta^{(2)}$. Here, we use the symbols $\mathcal{P}^{(2)}_1$, $\mathcal{P}^{(2)}_2$, and $\mathcal{P}^{(2)}_3$ to represent the contributions of the source terms $\mathcal{S}_1\sim \phi\phi$, $\mathcal{S}_2\sim \phi h$ and $\mathcal{S}_3\sim hh$ respectively. As shown in Fig.~\ref{fig:P_1}, for tensor-to-scalar ratio $r=A_{h}/A_{\zeta}=1$, the second order perturbations sourced by $\mathcal{S}_1\sim \phi\phi$ dominate the total power spectra of $\Phi^{(2)}$, $\Psi^{(2)}$, and $\delta^{(2)}$. 

\begin{figure}
	\includegraphics[scale = 0.33]{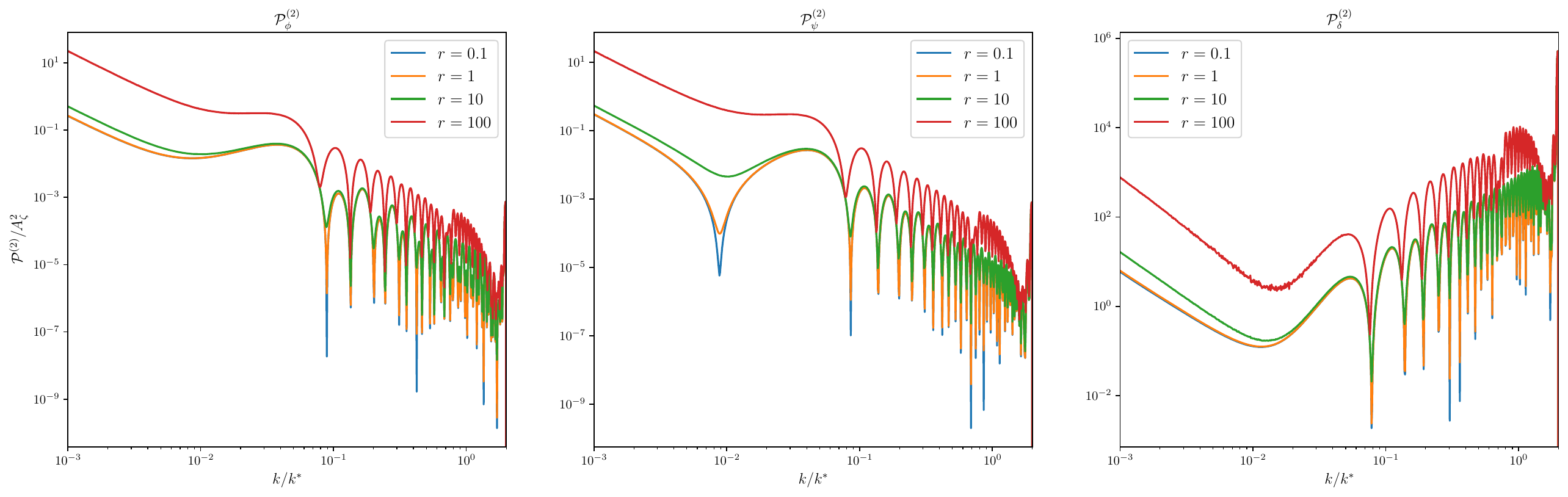}
	\caption{The total power spectra of second order perturbations $\Phi^{(2)}$, $\Psi^{(2)}$, and $\delta^{(2)}$ for different tensor-to-scalar ratio $r=A_{h}/A_{\zeta}$. We have set  $x_*=k_*\eta=100$. }\label{fig:ps}
\end{figure}

In order to study the effects of the large tensor-to-scalar ratio $r=A_{h}/A_{\zeta}$ on small scales, we calculate the total power spectra of  $\Phi^{(2)}$, $\Psi^{(2)}$, and $\delta^{(2)}$ for different $r=A_{h}/A_{\zeta}$. In Fig.~\ref{fig:ps}, the total power spectra of second order scalar perturbations for different $r$ are presented. For tensor-to-scalar ratio $r\ll 1$, the effects of primordial tensor perturbation become negligible, the total power spectra $\mathcal{P}^{(2)}_{\phi}$, $\mathcal{P}^{(2)}_{\psi}$, and $\mathcal{P}^{(2)}_{\delta}$ reduce to the results in Ref.~\cite{Inomata:2020cck}. For $r\gg 1$ on small scales, the effects of the primordial tensor perturbation become obvious, the total power spectra reduce to the results in Ref.~\cite{Bari:2021xvf,Nakama:2016enz,Cho:2020zbh}.


\subsection{Monochromatic primordial power spectra with different $k_*$}\label{sec:4.2}
\begin{figure}
	\includegraphics[scale = 0.33]{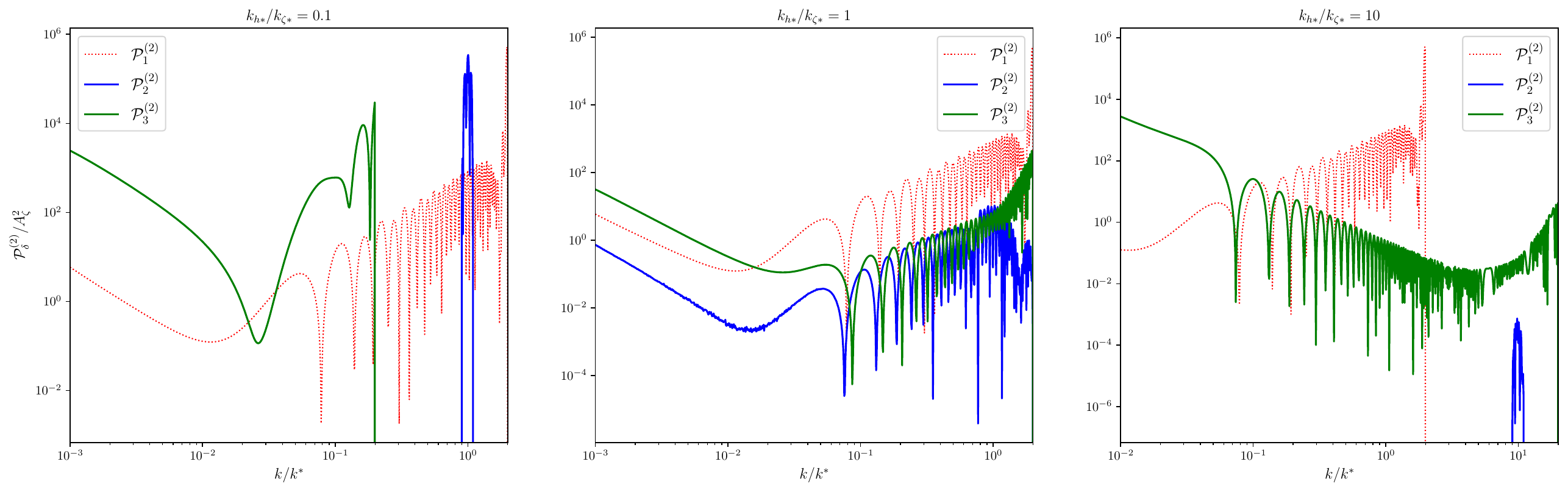}
	\caption{The power spectra $\mathcal{P}^{(2)}_a$$(a=1,2,3)$ for second order energy density perturbation $\delta^{(2)}$ with $k_{h *}/k_{\zeta *}=0.1$, $k_{h *}/k_{\zeta *}=1$, and $k_{h *}/k_{\zeta *}=10$ respectively. We have set the tensor-to-scalar ratio $r=A_{h}/A_{\zeta}=10$, $k_{\zeta *}=k_*$, and $x_*=k_*\eta=100$. }\label{fig:kh}
\end{figure}

The monochromatic primordial power spectra with different $k_*$ can be written as 
\begin{equation}
	\begin{aligned}
		\mathcal{P}_{\zeta}=A_{\zeta}k_{\zeta *}\delta\left( k-k_{\zeta *} \right) \ , \ \mathcal{P}_{h}=A_{h}k_{h *}\delta\left( k-k_{h*} \right) \ , \  k_{\zeta *}=k_*\ne k_{h *} \  .
	\end{aligned}
\end{equation}
In Fig.~\ref{fig:kh}, we plot the three kinds of power spectra $\mathcal{P}^{(2)}_a$$(a=1,2,3)$ in Eq.~(\ref{eq:p1})--Eq.~(\ref{eq:p3}) for second order energy density perturbation $\delta^{(2)}=\delta \rho^{(2)}/\rho^{(0)}$ with different $k_{h *}$. 
As shown in Fig.~\ref{fig:kh}, for $k_{h *}\ne k_{\zeta *}$, the behaviors of the power spectra $\mathcal{P}^{(2)}_2$ sourced by $\mathcal{S}_2\sim \phi h$ are different from the case of $k_{h *}/k_{\zeta *}=1$. More precisely, for $k_{h *}/k_{\zeta *}=n>1$, the domain of definition of $\mathcal{P}^{(2)}_2$ is $[(n-1)k/k_*,(n+1)k/k_*]$. For $k_{h *}/k_{\zeta *}=n\gg 1$, the power spectrum $\mathcal{P}^{(2)}_2$ becomes a small peak near $n\times k/k_*$. In this case, the contributions of the power spectra $\mathcal{P}^{(2)}_2$ can be neglected. And for $k_{h *}/k_{\zeta *}=n<1$, the domain of definition of $\mathcal{P}^{(2)}_2$ is $[(1-n)k/k_*,(1+n)k/k_*]$. For $k_{h *}/k_{\zeta *}=n\ll 1$, the power spectra $\mathcal{P}^{(2)}_2$ sourced by $\mathcal{S}_2\sim \phi h$ becomes a large peak near $k/k_*$.

\begin{figure}
	\includegraphics[scale = 0.33]{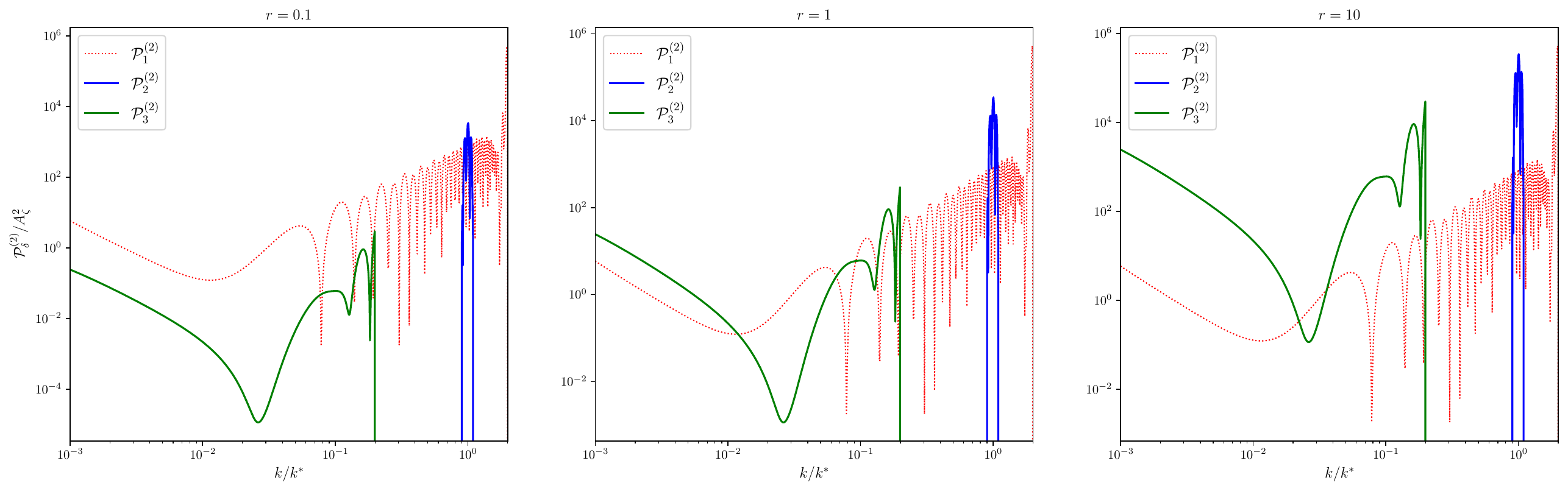}
	\caption{The power spectra $\mathcal{P}^{(2)}_a$$(a=1,2,3)$ for second order energy density perturbation $\delta^{(2)}$ with $r=0.1$, $r=1$, and $r=10$ respectively. We have set $n=k_{h*}/k_{\zeta*}=0.1$, $k_{\zeta *}=k_*$, and $x_*=k_*\eta=100$. }\label{fig:kr}
\end{figure}

As shown in Fig.~\ref{fig:kr}, for $k_{h *}/k_{\zeta *}=0.1$, the power spectra $\mathcal{P}^{(2)}_2$ can be larger than $\mathcal{P}^{(2)}_1$ sourced by $\mathcal{S}_1\sim \phi \phi$ even in the case of $r=0.1$. Note that the contributions of the source term $\mathcal{S}_2\sim \phi h$ in Eq.~(\ref{eq:S2}) are completely neglected in previous studies \cite{Inomata:2020cck,Bari:2021xvf,Nakama:2016enz}. Here, we point out that the contributions of the source term $\mathcal{S}_2\sim \phi h$ are very important for the monochromatic primordial power spectra with $k_{h *}/k_{\zeta *}\ll 1$.

\section{Log-normal primordial power spectra}\label{sec:4.0}
\begin{figure}
	\includegraphics[scale = 1]{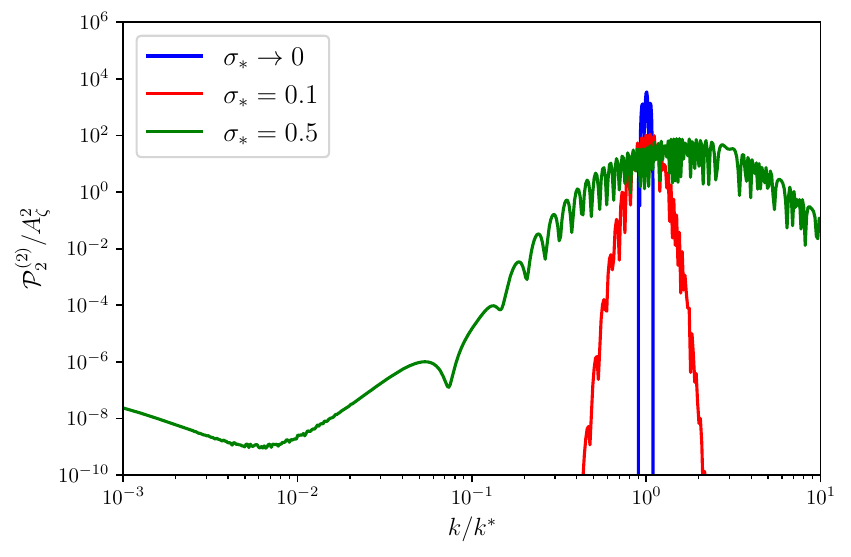}
	\caption{The power spectra $\mathcal{P}^{(2)}_2$ sourced by $\mathcal{S}_2\sim \phi h$ for second order energy density perturbation $\delta^{(2)}$ with $r=0.1$. We have set $n=k_{h*}/k_{\zeta*}=0.1$ and $x_*=k_*\eta=100$. }\label{fig:lp1}
\end{figure}

\begin{figure}
	\includegraphics[scale = 0.33]{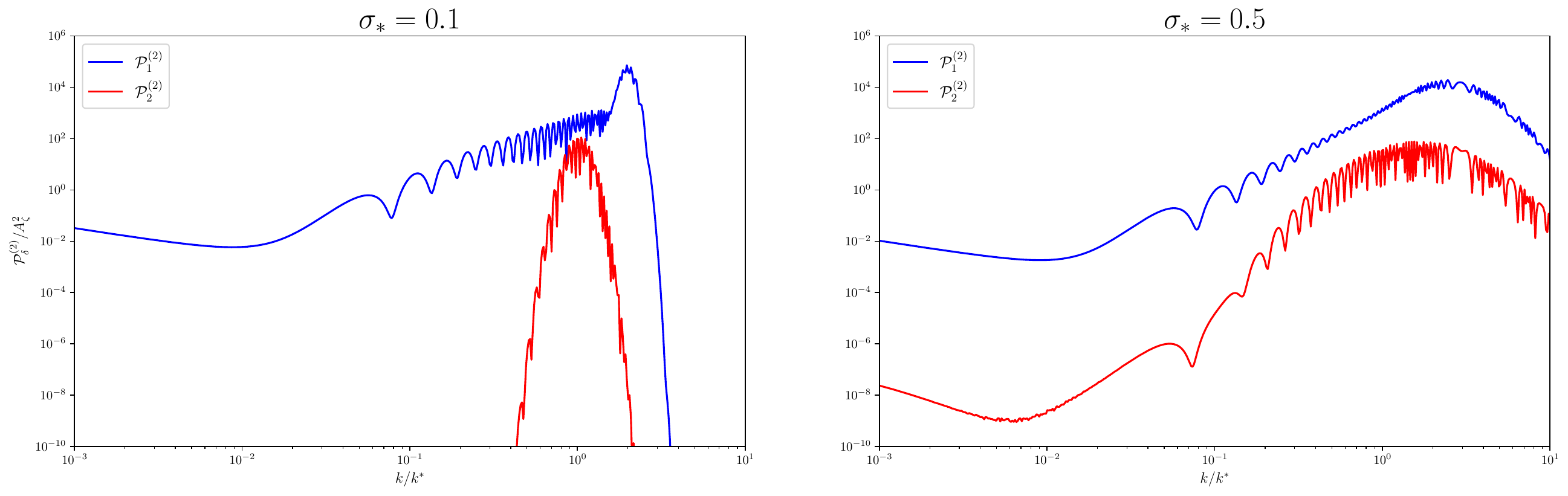}
	\caption{The power spectra $\mathcal{P}^{(2)}_1$ and $\mathcal{P}^{(2)}_2$ for second order energy density perturbation $\delta^{(2)}$ with $r=0.1$ for different $\sigma_*$. We have set $n=k_{h*}/k_{\zeta*}=0.1$ and $x_*=k_*\eta=100$. }\label{fig:lp2}
\end{figure}

Since the monochromatic primordial power spectra have infinitesimal width, it is necessary to consider a more realistic model, such as log-normal primordial power spectra. We consider the log-normal power spectra for primordial scalar and tensor perturbations. Here, we concentrated on the effects of the source term $\mathcal{S}_2\sim \phi h$ and corresponding second order power spectra $\mathcal{P}^{(2)}_2$. The log-normal primordial power spectra are given by
\begin{equation}
	\begin{aligned}
			\mathcal{P}_{\zeta}=\frac{A_{\zeta}}{\sqrt{2\pi \sigma^2_*}}\exp{\left(-\frac{\ln^2{\left(\frac{k}{k_{\zeta *}}\right)}}{2\sigma^2_*}\right)}  \ , \   	\mathcal{P}_{h}=\frac{A_{h}}{\sqrt{2\pi \sigma^2_*}}\exp{\left(-\frac{\ln^2{\left(\frac{k}{k_{h *}}\right)}}{2\sigma^2_*}\right)}  \  .
	\end{aligned}
\end{equation}
Here, we concentrate on the large peak of $\mathcal{P}^{(2)}_2$ with $k_{h *}/k_{\zeta *}\ll 1$. As mentioned in Sec.~\ref{sec:4.2}, for $k_{h *}/k_{\zeta *}=n\ll 1$, the power spectra $\mathcal{P}^{(2)}_2$ sourced by $\mathcal{S}_2\sim \phi h$ has a large peak near $k/k_*$. For the log-normal primordial power spectra, we calculate the power spectra of the second order density perturbation $\delta^{(2)}$ with $k_{h *}/k_{\zeta *}=0.1$. As shown in Fig.~\ref{fig:lp1}, for log-normal primordial power spectra, the peak of $\mathcal{P}^{(2)}_2$ becomes larger during the process of $\sigma_*\to 0$. 
For comparison, we plot $\mathcal{P}^{(2)}_1$ and $\mathcal{P}^{(2)}_2$ for the second order density perturbation $\delta^{(2)}$ with tensor-to-scalar ratio $r=0.1$ in Fig.~\ref{fig:lp2}. It shows that the contributions of $\mathcal{P}^{(2)}_2$ become smaller when $\sigma_*$ increases. Namely, the effects of the source term $\mathcal{S}_2\sim \phi h$ become more obviously when $\sigma_*$ and $n$ become smaller.
\section{ Conclusions and discussions}\label{sec:5}
In this paper, we systematically studied the second order density perturbations induced by primordial gravitational wave and primordial scalar perturbation. Since the constraints of primordial curvature and tensor perturbations on small scales are significantly weaker than those on large scales, we considered the large tensor-to-scalar ratio $r$ on small scales. As shown in Fig.~\ref{fig:ps}, the effects of the primordial tensor perturbation become obvious for $r\gg 1$. For tensor-to-scalar ratio $r\ll 1$, the effects of primordial tensor perturbation become negligible and our results of the power spectra $\mathcal{P}^{(2)}_{\phi}$, $\mathcal{P}^{(2)}_{\psi}$, and $\mathcal{P}^{(2)}_{\delta}$ reduce to the previous results in Ref.~\cite{Inomata:2020cck}. 

We give the explicit expressions for the power spectra of primordial scalar and tensor induced scalar and density perturbations in Eqs.~(\ref{eq:p1})--(\ref{eq:delta}). Specifically, for a given primordial scalar and tensor power spectra, the power spectra of the second order induced scalar and density perturbations can be calculated using these equations. In this paper, we considered the  primordial power spectra following delta and log-normal on small scales. It is essential to explore more general forms of the primordial power spectra, such as the log-normal primordial scalar power spectrum and the power-law primordial tensor power spectrum \cite{Yu:2023lmo}.

Moreover, the second order induced density perturbations offer a window to understand small-scale primordial gravitational waves and primordial curvature perturbations. More precisely, the primordial scalar and tensor power spectra can be calculated in terms of a given inflation model. Using Eqs.~(\ref{eq:p1})--(\ref{eq:delta}), we can calculate the power spectra of second order induced scalar and energy density perturbations. The induced density perturbations will affect many physical processes at small scales, such as the formation of PBH \cite{DeLuca:2023tun} and the high order GW background \cite{Chen:2022dah}. By observing PBHs or higher order GWs, we can constrain the power spectrum of second order induced density perturbations, thereby constraining inflationary models and the physical properties of primordial scalar and gravitational wave on small scales. Related research might be given in future work.

% We can also consider the primordial scalar and tensor power spectrum generated by a given inflation model in future work, and use this as a probe to test inflation models at small scales in the future.
%%%%%%%%%%%%%%%%%%%%%%%%%%%%%%%%%%%%%%%%%%
\vspace{6pt} 

\appendix
\section{Source terms}\label{sec:S} 
\begin{equation}
	\begin{aligned}
		\mathcal{S}_{ij,1}\left(\mathbf{x},\eta  \right)&=4\phi\partial_i\partial_j\phi+\partial_i\phi\partial_j\phi-\frac{\partial_i\phi'\partial_j\phi}{\mathcal{H}}-\frac{\partial_i\phi\partial_j\phi'}{\mathcal{H}}-\frac{\partial_i\phi'\partial_j\phi'}{\mathcal{H}^2} +\delta_{ij}\left(-24\mathcal{H}\phi\phi' \right. \\
		&\left.-2\left( \phi' \right)^2-4\phi\phi''-\frac{16}{3}\phi\Delta \phi-\frac{11}{3}\partial_b\phi\partial^b\phi+\frac{2}{3\mathcal{H}}\partial_b\phi'\partial^b\phi +\frac{1}{3\mathcal{H}^2}\partial_b\phi'\partial^b\phi' \right) \ ,
		\label{eq:S1}
	\end{aligned} 
\end{equation}
\begin{equation}
	\begin{aligned}
		\mathcal{S}_{ij,2}\left(\mathbf{x},\eta  \right)&=-h''_{ij}\phi-2\mathcal{H}h'_{ij}\phi+10\mathcal{H}h_{ij}\phi'+3h_{ij}\phi''-\phi\Delta h_{ij}-\frac{5}{3}h_{ij}\Delta \phi+h_{j}^{b}\partial_b\partial_i\phi  \\
		&+h^b_i\partial_b\partial_j\phi-2\partial_bh_{ij}\partial^b\phi+\partial^b\phi\partial_ih_{jb}+\partial^b\phi\partial_jh_{ib}-\delta_{ij}\frac{2}{3}h^{bc}\partial_c\partial_b\phi \ ,
		\label{eq:S2}
	\end{aligned} 
\end{equation}
\begin{equation}
	\begin{aligned}
		\mathcal{S}_{ij,3}\left(\mathbf{x},\eta  \right)&=-\frac{1}{2}h^{b'}_{i}h'_{jb}+\frac{1}{2}h^{bc}\partial_c\partial_b h_{ij}-\frac{1}{2}h^{bc}\partial_c\partial_i h_{jb}-\frac{1}{2}h^{bc}\partial_c\partial_jh_{ib}-\frac{1}{2}\partial_bh_{jc}\partial^ch^{b}_{i} \\
		&+\frac{1}{2}\partial_ch_{jb}\partial^ch^b_i+\frac{1}{4}\partial_ih^{bc}\partial_jh_{bc}+\frac{1}{2}h^{bc}\partial_i\partial_jh_{bc} +\delta_{ij}\left(\frac{5}{12}h'_{bc}h'^{bc}+\frac{1}{2}h^{bc}h''_{bc} \right. \\
		&\left.+\frac{4\mathcal{H}}{3}h^{bc}h'_{bc}-\frac{2}{3}h^{bc}\Delta h_{bc}+\frac{1}{3}\partial_c h_{bd}\partial^dh^{bc}-\frac{1}{2}\partial_dh_{bc} \partial^{d}h^{bc}\right) \ ,
		\label{eq:S3}
	\end{aligned} 
\end{equation}
where in Eqs.~(\ref{eq:S1})--(\ref{eq:S3}), we have defined $\partial_{\eta}\phi\equiv \phi'$. The source terms in momentum space are given by
\begin{equation}\label{eq:s1}
	\begin{aligned}
		S_{ij,1}\left(\mathbf{k},\eta  \right)
		&=-\int\frac{d^3p}{(2\pi)^{3/2}} \left(4p_ip_jT_{\phi}(ux)T_{\phi}(vx)+\left((k-p)_ip_j  \right)\left(T_{\phi}(ux)T_{\phi}(vx)-ux \frac{d}{d(ux)}T_{\phi}(ux)T_{\phi}(vx) \right.\right.  \\
		&  \left. -vx T_{\phi}(ux)\frac{d}{d(vx)}T_{\phi}(vx)-vux^2 \frac{d}{d(ux)}T_{\phi}(ux)\frac{d}{d(vx)}T_{\phi}(vx) \right)-\delta_{ij} \left(  -\frac{24vk^2}{x}T_{\phi}(ux)T'_{\phi}(vx) \right. \\
		&-2uvk^2T'_{\phi}(ux)T'_{\phi}(vx)-4v^2k^2T_{\phi}(ux)T''_{\phi}(vx)   +\frac{16v^2k^2}{3}T_{\phi}(ux)T_{\phi}(vx) \\
		&+\frac{11k^2(1-u^2-v^2)}{6}T_{\phi}(ux)T_{\phi}(vx)  -\frac{k^2(1-u^2-v^2)ux}{3}T'_{\phi}(ux)T_{\phi}(vx)  \\
		&\left.\left. -\frac{k^2(1-u^2-v^2)x^2uv}{6}T'_{\phi}(ux)T'_{\phi}(vx)\right)  \right) \frac{4}{9}\zeta_{\mathbf{k}-\mathbf{p}}\zeta_{\mathbf{p}} \ ,
	\end{aligned} 
\end{equation}
\begin{equation}\label{eq:s2}
	\begin{aligned}
		S_{ij,2}\left(\mathbf{k},\eta  \right)&=\int\frac{d^3p}{(2\pi)^{3/2}} \left(-\left( (k-p)^bp_i \right)\varepsilon^{\lambda}_{jb}\left(\mathbf{p}\right)T_{\phi}(ux)T_{h}(vx)-\left( (k-p)^bp_j \right)\varepsilon^{\lambda}_{ib}\left(\mathbf{p}\right)T_{\phi}(ux)T_{h}(vx) \right.  \\
		& \left. -\left( (k-p)_b(k-p)_i \right)\varepsilon^{\lambda,b}_{j}\left(\mathbf{p}\right)T_{\phi}(ux)T_{h}(vx)-\left( (k-p)_b(k-p)_j \right)\varepsilon^{\lambda,b}_{i}\left(\mathbf{p}\right)T_{\phi}(ux)T_{h}(vx) \right. \\
		& \left. +\varepsilon^{\lambda}_{ij}\left(\mathbf{p}\right)\left(-v^2k^2T_{\phi}(ux)\frac{d^2}{d(vx)^2}T_{h}(vx)-\frac{2vk^2}{x}T_{\phi}(ux)\frac{d}{d(vx)}T_{h}(vx) \right.\right. \\
		& \left.\left.+3u^2k^2\frac{d^2}{d(ux)^2}T_{\phi}(ux)T_{h}(vx) +v^2k^2T_{\phi}(ux)T_{h}(vx)+\frac{5u^2k^2}{3}T_{\phi}(ux)T_{h}(vx) \right.\right.  \\
		& \left. +k^2\left(1-v^2-u^2 \right)T_{\phi}(ux)T_{h}(vx)+\frac{10uk^2}{x}\frac{d}{d(ux)}T_{\phi}(ux)T_{h}(vx)  \right) \\
		&\left.+\delta_{ij} \varepsilon^{\lambda,bc}(\mathbf{p})(k-p)_b(k-p)_c \left(\frac{2}{3}T_{\phi}(ux)T_h(vx)  \right) \right) \frac{2}{3} \zeta_{\mathbf{k}-\mathbf{p}}\mathbf{h}^{\lambda}_{\mathbf{p}} \ , 
	\end{aligned} 
\end{equation}
\begin{equation}\label{eq:s3}
	\begin{aligned}
		S_{ij,3}\left(\mathbf{k},\eta  \right)&=\int\frac{d^3p}{(2\pi)^{3/2}} \frac{1}{2} \Bigg( \varepsilon^{\lambda_1,b}_{i}\left(\mathbf{k} - \mathbf{p}\right)\varepsilon^{\lambda_2}_{jb}\left(\mathbf{p}\right)\left(-uvk^2\frac{d}{(ux)}T_{h}(ux)\frac{d}{(vx)}T_h(vx) \right. \\
		& \left. -\frac{k^2(1-v^2-u^2)}{2}T_h(ux)T_h(vx) \right)-\varepsilon^{\lambda_1,bc}(\mathbf{k}-\mathbf{p})\varepsilon^{\lambda_2}_{ij}(\mathbf{p})p_bp_cT_{h}(ux)T_{h}(vx) \\
		&  +\varepsilon^{\lambda_1,bc}(\mathbf{k}-\mathbf{p})\varepsilon^{\lambda_2}_{jb}(\mathbf{p})p_cp_iT_{h}(ux)T_{h}(vx)+\varepsilon^{\lambda_1,bc}(\mathbf{k}-\mathbf{p})\varepsilon^{\lambda_2}_{ib}(\mathbf{p})p_cp_jT_{h}(ux)T_{h}(vx) \\
		&  -\varepsilon^{\lambda_1,bc}(\mathbf{k}-\mathbf{p})\varepsilon^{\lambda_2}_{bc}(\mathbf{p}) \left( \frac{1}{2}(k-p)_ip_jT_h(ux)T_h(vx)+p_ip_jT_h(ux)T_h(vx) \right) \\
		&  +\varepsilon^{\lambda_1}_{jc}(\mathbf{k}-\mathbf{p})\varepsilon^{\lambda_2,b}_{i}(\mathbf{p})(k-p)_bp^cT_{h}(ux)T_{h}(vx)\\
		&+2\delta_{ij} \left(\varepsilon^{\lambda_1}_{bc}(\mathbf{k}-\mathbf{p})\varepsilon^{\lambda_2,bc}(\mathbf{p}) \left( \frac{5uvk^2}{12}T'_h(ux)T'_h(vx)+\frac{v^2k^2}{2}T_h(ux)T''_h(vx)\right. \right. \\
		& \left. +\frac{4vk^2}{3x}T_h(ux)T'_h(vx)+\frac{2v^2k^2}{3}T_h(ux)T_h(vx)+\frac{k^2(1-u^2-v^2)}{4}T_h(ux)T_h(vx)\right) \\
		&  \left.-\frac{1}{3}\varepsilon^{\lambda_1}_{bd}(\mathbf{k}-\mathbf{p})\varepsilon^{\lambda_2,bc}(\mathbf{p})(k-p)_cp^dT_h(ux)T_h(vx) \right)
		\Bigg)\mathbf{h}^{\lambda_1}_{\mathbf{k}-\mathbf{p}}\mathbf{h}^{\lambda_2}_{\mathbf{p}}   \ .
	\end{aligned} 
\end{equation}
In Eqs.~(\ref{eq:s1})--(\ref{eq:s3}), we have defined $|\mathbf{k}-\mathbf{p}|=u|\mathbf{k}|=uk$ and $p=|\mathbf{p}|=vk$.The explicit expressions of the polarization tensor $\varepsilon^{\lambda_1,ij}(\mathbf{p})$ are given in~\ref{sec:A}. The first order scalar and tensor perturbations in Eqs.~(\ref{eq:s1})--(\ref{eq:s3}) have been written as
\begin{equation}
	\psi(\eta,\mathbf{k}) = \phi(\eta,\mathbf{k}) = \frac{2}{3}\zeta_{\mathbf{k}} T_\phi(k \eta) \ , \
	h^{\lambda}(\eta,\mathbf{k}) = \mathbf{h}^{\lambda}_{\mathbf{k}} T_{h}(k \eta) \ ,
\end{equation}
where $\zeta_{\mathbf{k}}$ and $\mathbf{h}_{\mathbf{k}}$ are the primordial curvature and tensor perturbations respectively. The transfer functions $ T_\phi(k \eta)$ and $T_{h}(k \eta)$ in the RD era are given by \cite{Inomata:2020cck}
\begin{equation}\label{eq:T}
	T_{\phi}(x)=\frac{9}{x^{2}}\left(\frac{\sqrt{3}}{x} \sin \left(\frac{x}{\sqrt{3}}\right)-\cos \left(\frac{x}{\sqrt{3}}\right)\right) \ , \  T_{h}=\frac{\sin x}{x} \ .
\end{equation} 
\section{Polarization tensor}\label{sec:A}
The polarization tensor is defined as
\begin{equation}
	\begin{aligned}
		\varepsilon^{\times}_{ij}\left(\mathbf{k}  \right)=\frac{1}{\sqrt{2}}\left( e_i\left( \mathbf{k} \right)\bar{e}_j\left( \mathbf{k} \right)+\bar{e}_i\left( \mathbf{k} \right)e_j\left( \mathbf{k} \right)  \right) \ ,
	\end{aligned} 
\end{equation}
\begin{equation}
	\begin{aligned}
		\varepsilon^{+}_{ij}\left(\mathbf{k}  \right)=\frac{1}{\sqrt{2}}\left( e_i\left( \mathbf{k} \right)e_j\left( \mathbf{k} \right)-\bar{e}_i\left( \mathbf{k} \right)\bar{e}_j\left( \mathbf{k} \right)  \right) \ ,
	\end{aligned} 
\end{equation}
where $\left(\mathbf{k}_i/|k|,e_i\left( \mathbf{k} \right),\bar{e}_i\left( \mathbf{k} \right)  \right)$ is the normalized bases in three dimensional momentum space. We study the polarization tensor for a given coordinate system, namely
\begin{equation}
	\begin{aligned}
		\mathbf{k}=\left(0,0,k  \right) \ , \ e_i\left( \mathbf{k} \right)=\left( 1,0,0 \right) \ , \ \bar{e}_i\left( \mathbf{k} \right)=\left( 0,1,0 \right) \ .
	\end{aligned} 
\end{equation}
Then the polarization tensors $	\varepsilon^{\lambda}_{ij}\left(\mathbf{k}-\mathbf{p}  \right)$ and $	\varepsilon^{\lambda}_{ij}\left(\mathbf{p}  \right)$ can be written as
\begin{equation}
	\begin{aligned}
		\varepsilon^{\times}_{ij}\left(\mathbf{k}-\mathbf{p}  \right)&=\frac{1}{\sqrt{2}}\left( e_i\left( \mathbf{k}-\mathbf{p} \right)\bar{e}_j\left( \mathbf{k}-\mathbf{p} \right)+\bar{e}_i\left( \mathbf{k}-\mathbf{p} \right)e_j\left( \mathbf{k}-\mathbf{p} \right)  \right) \ , \\
		\varepsilon^{+}_{ij}\left(\mathbf{k}-\mathbf{p}  \right)&=\frac{1}{\sqrt{2}}\left( e_i\left( \mathbf{k}-\mathbf{p} \right)e_j\left( \mathbf{k}-\mathbf{p} \right)-\bar{e}_i\left( \mathbf{k}-\mathbf{p} \right)\bar{e}_j\left( \mathbf{k}-\mathbf{p} \right)  \right) \ , \\
		\varepsilon^{\times}_{ij}\left(\mathbf{p}  \right)&=\frac{1}{\sqrt{2}}\left( e_i\left( \mathbf{p} \right)\bar{e}_j\left( \mathbf{p} \right)+\bar{e}_i\left( \mathbf{p} \right)e_j\left( \mathbf{p} \right)  \right)\ , \\ \varepsilon^{+}_{ij}\left(\mathbf{p}  \right)&=\frac{1}{\sqrt{2}}\left( e_i\left( \mathbf{p} \right)e_j\left( \mathbf{p} \right)-\bar{e}_i\left( \mathbf{p} \right)\bar{e}_j\left( \mathbf{p} \right)  \right) \ ,
	\end{aligned} 
\end{equation}
where 
\begin{equation}
	\begin{aligned}
		\mathbf{k}-\mathbf{p}&=k\left(-\sqrt{ v^2-\frac{1}{4}  \left(-u^2+v^2+1\right)^2},0,\frac{1}{2} \left(u^2-v^2+1\right)  \right)  \ , \\
		e_i\left( \mathbf{k}-\mathbf{p} \right)&=\left(\frac{u^2-v^2+1}{2 u},0,\frac{\sqrt{-u^4+2 u^2 v^2+2 u^2-v^4+2 v^2-1}}{2 u}  \right)\ , \\
		\bar{e}_i\left( \mathbf{k}-\mathbf{p} \right)&=\left( 0,1,0 \right) \ ,
	\end{aligned} 
\end{equation}
\begin{equation}
	\begin{aligned}
		\mathbf{p}&=k\left(\sqrt{ v^2-\frac{1}{4} \left(-u^2+v^2+1\right)^2},0,\frac{1}{2}  \left(-u^2+v^2+1\right) \right)  \ , \\
		e_i\left( \mathbf{p} \right)&=\left(-\frac{-u^2+v^2+1}{2 v},0,\frac{\sqrt{-u^4+2 u^2 \left(v^2+1\right)-\left(v^2-1\right)^2}}{2 v}  \right)\ , \\
		\bar{e}_i\left( \mathbf{p} \right)&=\left( 0,1,0 \right) \ .
	\end{aligned} 
\end{equation}
\section{four-point function}\label{sec:B}
	\begin{equation}\label{eq:B1}
		\begin{aligned}
		\left\langle\zeta_{\mathbf{k}-\mathbf{p}} \zeta_{\mathbf{p}} \zeta_{\mathbf{k}'-\mathbf{p}'} \zeta_{\mathbf{p}'}\right\rangle&=\left\langle\zeta_{\mathbf{k}-\mathbf{p}} \zeta_{\mathbf{k}'-\mathbf{p}'}\right\rangle\left\langle\zeta_{\mathbf{p}} \zeta_{\mathbf{p}'}\right\rangle+\left\langle\zeta_{\mathbf{k}-\mathbf{p}} \zeta_{\mathbf{p}'}\right\rangle\left\langle\zeta_{\mathbf{p}} \zeta_{\mathbf{k}'-\mathbf{p}'}\right\rangle \\
		&=\delta\left(\mathbf{k}+\mathbf{k}'\right) \frac{(2\pi^2)^2}{p^3|\mathbf{k}-\mathbf{p}|^3} \left(\delta\left(\mathbf{p}+\mathbf{p}'\right)+\delta\left(\mathbf{k}'-\mathbf{p}'+\mathbf{p}\right) \right) \mathcal{P}_{\zeta}(|\mathbf{k}-\mathbf{p}|)\mathcal{P}_{\zeta}(p) \ ,
			\end{aligned}
	\end{equation}

\begin{equation}\label{eq:B2}
	\begin{aligned}
		\langle \zeta_{\mathbf{k}-\mathbf{p}}\mathbf{h}_{\mathbf{p}}^{\lambda_1} \zeta_{\mathbf{k}'-\mathbf{p}'}\mathbf{h}_{\mathbf{p}'}^{\lambda'_1}  \rangle =&\langle \zeta_{\mathbf{k}-\mathbf{p}} \zeta_{\mathbf{k}'-\mathbf{p}'}\rangle\langle \mathbf{h}_{\mathbf{p}}^{\lambda_1}\mathbf{h}_{\mathbf{p}'}^{\lambda'_1}  \rangle \\
		=&\delta\left(\mathbf{k}+\mathbf{k}'\right)\delta^{\lambda_1\lambda'_1} \frac{(2\pi^2)^2}{p^3|\mathbf{k}-\mathbf{p}|^3} \delta\left(\mathbf{p}+\mathbf{p}'\right)\mathcal{P}_{\zeta}(|\mathbf{k}-\mathbf{p}|)\mathcal{P}_{h}(p) \ ,
	\end{aligned}
\end{equation}

\begin{equation}\label{eq:B3}
	\begin{aligned}
		\langle \mathbf{h}_{\mathbf{k}-\mathbf{p}}^{\lambda_1}\mathbf{h}_{\mathbf{p}}^{\lambda_2} \mathbf{h}_{\mathbf{k}'-\mathbf{p}'}^{\lambda_1'}\mathbf{h}_{\mathbf{p}'}^{\lambda_2'} \rangle
		&=\delta\left(\mathbf{k}+\mathbf{k}'\right) \frac{(2\pi^2)^2}{p^3|\mathbf{k}-\mathbf{p}|^3} \left(\delta^{\lambda_1\lambda'_1}\delta^{\lambda_2\lambda'_2}\delta\left(\mathbf{p}+\mathbf{p}'\right)+\delta^{\lambda_1\lambda'_2}\delta^{\lambda_2\lambda'_1}\delta\left(\mathbf{k}'-\mathbf{p}'+\mathbf{p}\right) \right)\\
		&\times \mathcal{P}_{h}(|\mathbf{k}-\mathbf{p}|)\mathcal{P}_{h}(p) \ .
	\end{aligned}
\end{equation}

\bibliography{biblio}

\begin{thebibliography}{999}

\bibitem[Aghanim et~al.(2020)]{Planck:2018vyg}
Aghanim, N.;  et~al.
\newblock {Planck 2018 results. VI. Cosmological parameters}.
\newblock {\em Astron. Astrophys.} {\bf 2020}, {\em 641},~A6,
  \href{http://xxx.lanl.gov/abs/1807.06209}{{\normalfont
  [arXiv:astro-ph.CO/1807.06209]}}.
\newblock [Erratum: Astron.Astrophys. 652, C4 (2021)],
  {\url{https://doi.org/10.1051/0004-6361/201833910}}.

\bibitem[Malik and Wands(2009)]{Malik:2008im}
Malik, K.A.; Wands, D.
\newblock {Cosmological perturbations}.
\newblock {\em Phys. Rept.} {\bf 2009}, {\em 475},~1--51,
  \href{http://xxx.lanl.gov/abs/0809.4944}{{\normalfont
  [arXiv:astro-ph/0809.4944]}}.
\newblock {\url{https://doi.org/10.1016/j.physrep.2009.03.001}}.

\bibitem[Mukhanov et~al.(1992)Mukhanov, Feldman, and
  Brandenberger]{Mukhanov:1990me}
Mukhanov, V.F.; Feldman, H.A.; Brandenberger, R.H.
\newblock {Theory of cosmological perturbations. Part 1. Classical
  perturbations. Part 2. Quantum theory of perturbations. Part 3. Extensions}.
\newblock {\em Phys. Rept.} {\bf 1992}, {\em 215},~203--333.
\newblock {\url{https://doi.org/10.1016/0370-1573(92)90044-Z}}.

\bibitem[Kodama and Sasaki(1984)]{Kodama:1984ziu}
Kodama, H.; Sasaki, M.
\newblock {Cosmological Perturbation Theory}.
\newblock {\em Prog. Theor. Phys. Suppl.} {\bf 1984}, {\em 78},~1--166.
\newblock {\url{https://doi.org/10.1143/PTPS.78.1}}.

\bibitem[Mollerach et~al.(2004)Mollerach, Harari, and
  Matarrese]{Mollerach:2003nq}
Mollerach, S.; Harari, D.; Matarrese, S.
\newblock {CMB polarization from secondary vector and tensor modes}.
\newblock {\em Phys. Rev. D} {\bf 2004}, {\em 69},~063002,
  \href{http://xxx.lanl.gov/abs/astro-ph/0310711}{{\normalfont
  [astro-ph/0310711]}}.
\newblock {\url{https://doi.org/10.1103/PhysRevD.69.063002}}.

\bibitem[Ananda et~al.(2007)Ananda, Clarkson, and Wands]{Ananda:2006af}
Ananda, K.N.; Clarkson, C.; Wands, D.
\newblock {The Cosmological gravitational wave background from primordial
  density perturbations}.
\newblock {\em Phys. Rev. D} {\bf 2007}, {\em 75},~123518,
  \href{http://xxx.lanl.gov/abs/gr-qc/0612013}{{\normalfont [gr-qc/0612013]}}.
\newblock {\url{https://doi.org/10.1103/PhysRevD.75.123518}}.

\bibitem[Baumann et~al.(2007)Baumann, Steinhardt, Takahashi, and
  Ichiki]{Baumann:2007zm}
Baumann, D.; Steinhardt, P.J.; Takahashi, K.; Ichiki, K.
\newblock {Gravitational Wave Spectrum Induced by Primordial Scalar
  Perturbations}.
\newblock {\em Phys. Rev. D} {\bf 2007}, {\em 76},~084019,
  \href{http://xxx.lanl.gov/abs/hep-th/0703290}{{\normalfont
  [hep-th/0703290]}}.
\newblock {\url{https://doi.org/10.1103/PhysRevD.76.084019}}.

\bibitem[Kohri and Terada(2018)]{Kohri:2018awv}
Kohri, K.; Terada, T.
\newblock {Semianalytic calculation of gravitational wave spectrum nonlinearly
  induced from primordial curvature perturbations}.
\newblock {\em Phys. Rev. D} {\bf 2018}, {\em 97},~123532,
  \href{http://xxx.lanl.gov/abs/1804.08577}{{\normalfont
  [arXiv:gr-qc/1804.08577]}}.
\newblock {\url{https://doi.org/10.1103/PhysRevD.97.123532}}.

\bibitem[Dom\`enech(2021)]{Domenech:2021ztg}
Dom\`enech, G.
\newblock {Scalar Induced Gravitational Waves Review}.
\newblock {\em Universe} {\bf 2021}, {\em 7},~398,
  \href{http://xxx.lanl.gov/abs/2109.01398}{{\normalfont
  [arXiv:gr-qc/2109.01398]}}.
\newblock {\url{https://doi.org/10.3390/universe7110398}}.

\bibitem[Chang et~al.(2022)Chang, Zhang, and Zhou]{Chang:2022nzu}
Chang, Z.; Zhang, X.; Zhou, J.Z.
\newblock {Primordial black holes and third order scalar induced gravitational
  waves} {\bf 2022}.
\newblock  \href{http://xxx.lanl.gov/abs/2209.12404}{{\normalfont
  [arXiv:astro-ph.CO/2209.12404]}}.

\bibitem[Romero-Rodriguez et~al.(2022)Romero-Rodriguez, Martinez, Pujol\`as,
  Sakellariadou, and Vaskonen]{Romero-Rodriguez:2021aws}
Romero-Rodriguez, A.; Martinez, M.; Pujol\`as, O.; Sakellariadou, M.; Vaskonen,
  V.
\newblock {Search for a Scalar Induced Stochastic Gravitational Wave Background
  in the Third LIGO-Virgo Observing Run}.
\newblock {\em Phys. Rev. Lett.} {\bf 2022}, {\em 128},~051301,
  \href{http://xxx.lanl.gov/abs/2107.11660}{{\normalfont
  [arXiv:gr-qc/2107.11660]}}.
\newblock {\url{https://doi.org/10.1103/PhysRevLett.128.051301}}.

\bibitem[Saito and Yokoyama(2009)]{Saito:2008jc}
Saito, R.; Yokoyama, J.
\newblock {Gravitational wave background as a probe of the primordial black
  hole abundance}.
\newblock {\em Phys. Rev. Lett.} {\bf 2009}, {\em 102},~161101,
  \href{http://xxx.lanl.gov/abs/0812.4339}{{\normalfont
  [arXiv:astro-ph/0812.4339]}}.
\newblock [Erratum: Phys.Rev.Lett. 107, 069901 (2011)],
  {\url{https://doi.org/10.1103/PhysRevLett.102.161101}}.

\bibitem[Wang et~al.(2019)Wang, Terada, and Kohri]{Wang:2019kaf}
Wang, S.; Terada, T.; Kohri, K.
\newblock {Prospective constraints on the primordial black hole abundance from
  the stochastic gravitational-wave backgrounds produced by coalescing events
  and curvature perturbations}.
\newblock {\em Phys. Rev. D} {\bf 2019}, {\em 99},~103531,
  \href{http://xxx.lanl.gov/abs/1903.05924}{{\normalfont
  [arXiv:astro-ph.CO/1903.05924]}}.
\newblock [Erratum: Phys.Rev.D 101, 069901 (2020)],
  {\url{https://doi.org/10.1103/PhysRevD.99.103531}}.

\bibitem[Inomata and Nakama(2019)]{Inomata:2018epa}
Inomata, K.; Nakama, T.
\newblock {Gravitational waves induced by scalar perturbations as probes of the
  small-scale primordial spectrum}.
\newblock {\em Phys. Rev. D} {\bf 2019}, {\em 99},~043511,
  \href{http://xxx.lanl.gov/abs/1812.00674}{{\normalfont
  [arXiv:astro-ph.CO/1812.00674]}}.
\newblock {\url{https://doi.org/10.1103/PhysRevD.99.043511}}.

\bibitem[Barausse et~al.(2020)]{Barausse:2020rsu}
Barausse, E.;  et~al.
\newblock {Prospects for Fundamental Physics with LISA}.
\newblock {\em Gen. Rel. Grav.} {\bf 2020}, {\em 52},~81,
  \href{http://xxx.lanl.gov/abs/2001.09793}{{\normalfont
  [arXiv:gr-qc/2001.09793]}}.
\newblock {\url{https://doi.org/10.1007/s10714-020-02691-1}}.

\bibitem[Bartolo et~al.(2019)Bartolo, De~Luca, Franciolini, Lewis, Peloso, and
  Riotto]{Bartolo:2018evs}
Bartolo, N.; De~Luca, V.; Franciolini, G.; Lewis, A.; Peloso, M.; Riotto, A.
\newblock {Primordial Black Hole Dark Matter: LISA Serendipity}.
\newblock {\em Phys. Rev. Lett.} {\bf 2019}, {\em 122},~211301,
  \href{http://xxx.lanl.gov/abs/1810.12218}{{\normalfont
  [arXiv:astro-ph.CO/1810.12218]}}.
\newblock {\url{https://doi.org/10.1103/PhysRevLett.122.211301}}.

\bibitem[Cai et~al.(2019)Cai, Pi, and Sasaki]{Cai:2018dig}
Cai, R.g.; Pi, S.; Sasaki, M.
\newblock {Gravitational Waves Induced by non-Gaussian Scalar Perturbations}.
\newblock {\em Phys. Rev. Lett.} {\bf 2019}, {\em 122},~201101,
  \href{http://xxx.lanl.gov/abs/1810.11000}{{\normalfont
  [arXiv:astro-ph.CO/1810.11000]}}.
\newblock {\url{https://doi.org/10.1103/PhysRevLett.122.201101}}.

\bibitem[Garcia-Bellido et~al.(2017)Garcia-Bellido, Peloso, and
  Unal]{Garcia-Bellido:2017aan}
Garcia-Bellido, J.; Peloso, M.; Unal, C.
\newblock {Gravitational Wave signatures of inflationary models from Primordial
  Black Hole Dark Matter}.
\newblock {\em JCAP} {\bf 2017}, {\em 09},~013,
  \href{http://xxx.lanl.gov/abs/1707.02441}{{\normalfont
  [arXiv:astro-ph.CO/1707.02441]}}.
\newblock {\url{https://doi.org/10.1088/1475-7516/2017/09/013}}.

\bibitem[Chang et~al.(2023)Chang, Zhang, and Zhou]{Chang:2022vlv}
Chang, Z.; Zhang, X.; Zhou, J.Z.
\newblock {Gravitational waves from primordial scalar and tensor
  perturbations}.
\newblock {\em Phys. Rev. D} {\bf 2023}, {\em 107},~063510,
  \href{http://xxx.lanl.gov/abs/2209.07693}{{\normalfont
  [arXiv:astro-ph.CO/2209.07693]}}.
\newblock {\url{https://doi.org/10.1103/PhysRevD.107.063510}}.

\bibitem[Chen et~al.(2022)Chen, Ota, Zhu, and Zhu]{Chen:2022dah}
Chen, C.; Ota, A.; Zhu, H.Y.; Zhu, Y.
\newblock {Missing one-loop contributions in secondary gravitational waves}
  {\bf 2022}.
\newblock  \href{http://xxx.lanl.gov/abs/2210.17176}{{\normalfont
  [arXiv:astro-ph.CO/2210.17176]}}.

\bibitem[Inomata et~al.(2017)Inomata, Kawasaki, Mukaida, Tada, and
  Yanagida]{Inomata:2016rbd}
Inomata, K.; Kawasaki, M.; Mukaida, K.; Tada, Y.; Yanagida, T.T.
\newblock {Inflationary primordial black holes for the LIGO gravitational wave
  events and pulsar timing array experiments}.
\newblock {\em Phys. Rev. D} {\bf 2017}, {\em 95},~123510,
  \href{http://xxx.lanl.gov/abs/1611.06130}{{\normalfont
  [arXiv:astro-ph.CO/1611.06130]}}.
\newblock {\url{https://doi.org/10.1103/PhysRevD.95.123510}}.

\bibitem[Alabidi et~al.(2012)Alabidi, Kohri, Sasaki, and
  Sendouda]{Alabidi:2012ex}
Alabidi, L.; Kohri, K.; Sasaki, M.; Sendouda, Y.
\newblock {Observable Spectra of Induced Gravitational Waves from Inflation}.
\newblock {\em JCAP} {\bf 2012}, {\em 09},~017,
  \href{http://xxx.lanl.gov/abs/1203.4663}{{\normalfont
  [arXiv:astro-ph.CO/1203.4663]}}.
\newblock {\url{https://doi.org/10.1088/1475-7516/2012/09/017}}.

\bibitem[Bugaev and Klimai(2010)]{Bugaev:2009zh}
Bugaev, E.; Klimai, P.
\newblock {Induced gravitational wave background and primordial black holes}.
\newblock {\em Phys. Rev. D} {\bf 2010}, {\em 81},~023517,
  \href{http://xxx.lanl.gov/abs/0908.0664}{{\normalfont
  [arXiv:astro-ph.CO/0908.0664]}}.
\newblock {\url{https://doi.org/10.1103/PhysRevD.81.023517}}.

\bibitem[Papanikolaou et~al.(2021)Papanikolaou, Vennin, and
  Langlois]{Papanikolaou:2020qtd}
Papanikolaou, T.; Vennin, V.; Langlois, D.
\newblock {Gravitational waves from a universe filled with primordial black
  holes}.
\newblock {\em JCAP} {\bf 2021}, {\em 03},~053,
  \href{http://xxx.lanl.gov/abs/2010.11573}{{\normalfont
  [arXiv:astro-ph.CO/2010.11573]}}.
\newblock {\url{https://doi.org/10.1088/1475-7516/2021/03/053}}.

\bibitem[Zhou et~al.(2020)Zhou, Jiang, Cai, Sasaki, and Pi]{Zhou:2020kkf}
Zhou, Z.; Jiang, J.; Cai, Y.F.; Sasaki, M.; Pi, S.
\newblock {Primordial black holes and gravitational waves from resonant
  amplification during inflation}.
\newblock {\em Phys. Rev. D} {\bf 2020}, {\em 102},~103527,
  \href{http://xxx.lanl.gov/abs/2010.03537}{{\normalfont
  [arXiv:astro-ph.CO/2010.03537]}}.
\newblock {\url{https://doi.org/10.1103/PhysRevD.102.103527}}.

\bibitem[Dom\`enech et~al.(2020)Dom\`enech, Pi, and Sasaki]{Domenech:2020kqm}
Dom\`enech, G.; Pi, S.; Sasaki, M.
\newblock {Induced gravitational waves as a probe of thermal history of the
  universe}.
\newblock {\em JCAP} {\bf 2020}, {\em 08},~017,
  \href{http://xxx.lanl.gov/abs/2005.12314}{{\normalfont
  [arXiv:gr-qc/2005.12314]}}.
\newblock {\url{https://doi.org/10.1088/1475-7516/2020/08/017}}.

\bibitem[Cai et~al.(2020)Cai, Guo, Liu, Liu, and Yang]{Cai:2019bmk}
Cai, R.G.; Guo, Z.K.; Liu, J.; Liu, L.; Yang, X.Y.
\newblock {Primordial black holes and gravitational waves from parametric
  amplification of curvature perturbations}.
\newblock {\em JCAP} {\bf 2020}, {\em 06},~013,
  \href{http://xxx.lanl.gov/abs/1912.10437}{{\normalfont
  [arXiv:astro-ph.CO/1912.10437]}}.
\newblock {\url{https://doi.org/10.1088/1475-7516/2020/06/013}}.

\bibitem[Cai et~al.(2019)Cai, Pi, Wang, and Yang]{Cai:2019elf}
Cai, R.G.; Pi, S.; Wang, S.J.; Yang, X.Y.
\newblock {Pulsar Timing Array Constraints on the Induced Gravitational Waves}.
\newblock {\em JCAP} {\bf 2019}, {\em 10},~059,
  \href{http://xxx.lanl.gov/abs/1907.06372}{{\normalfont
  [arXiv:astro-ph.CO/1907.06372]}}.
\newblock {\url{https://doi.org/10.1088/1475-7516/2019/10/059}}.

\bibitem[Yuan et~al.(2019)Yuan, Chen, and Huang]{Yuan:2019udt}
Yuan, C.; Chen, Z.C.; Huang, Q.G.
\newblock {Probing primordial\textendash{}black-hole dark matter with scalar
  induced gravitational waves}.
\newblock {\em Phys. Rev. D} {\bf 2019}, {\em 100},~081301,
  \href{http://xxx.lanl.gov/abs/1906.11549}{{\normalfont
  [arXiv:astro-ph.CO/1906.11549]}}.
\newblock {\url{https://doi.org/10.1103/PhysRevD.100.081301}}.

\bibitem[Bartolo et~al.(2018)Bartolo, Domcke, Figueroa, Garc\'\i{}a-Bellido,
  Peloso, Pieroni, Ricciardone, Sakellariadou, Sorbo, and
  Tasinato]{Bartolo:2018qqn}
Bartolo, N.; Domcke, V.; Figueroa, D.G.; Garc\'\i{}a-Bellido, J.; Peloso, M.;
  Pieroni, M.; Ricciardone, A.; Sakellariadou, M.; Sorbo, L.; Tasinato, G.
\newblock {Probing non-Gaussian Stochastic Gravitational Wave Backgrounds with
  LISA}.
\newblock {\em JCAP} {\bf 2018}, {\em 11},~034,
  \href{http://xxx.lanl.gov/abs/1806.02819}{{\normalfont
  [arXiv:astro-ph.CO/1806.02819]}}.
\newblock {\url{https://doi.org/10.1088/1475-7516/2018/11/034}}.

\bibitem[Alabidi et~al.(2013)Alabidi, Kohri, Sasaki, and
  Sendouda]{Alabidi:2013lya}
Alabidi, L.; Kohri, K.; Sasaki, M.; Sendouda, Y.
\newblock {Observable induced gravitational waves from an early matter phase}.
\newblock {\em JCAP} {\bf 2013}, {\em 05},~033,
  \href{http://xxx.lanl.gov/abs/1303.4519}{{\normalfont
  [arXiv:astro-ph.CO/1303.4519]}}.
\newblock {\url{https://doi.org/10.1088/1475-7516/2013/05/033}}.

\bibitem[Zhang et~al.(2022)Zhang, Zhou, and Chang]{Zhang:2022dgx}
Zhang, X.; Zhou, J.Z.; Chang, Z.
\newblock {Impact of the free-streaming neutrinos to the second order induced
  gravitational waves}.
\newblock {\em Eur. Phys. J. C} {\bf 2022}, {\em 82},~781,
  \href{http://xxx.lanl.gov/abs/2208.12948}{{\normalfont
  [arXiv:astro-ph.CO/2208.12948]}}.
\newblock {\url{https://doi.org/10.1140/epjc/s10052-022-10742-x}}.

\bibitem[Gong(2022)]{Gong:2019mui}
Gong, J.O.
\newblock {Analytic Integral Solutions for Induced Gravitational Waves}.
\newblock {\em Astrophys. J.} {\bf 2022}, {\em 925},~102,
  \href{http://xxx.lanl.gov/abs/1909.12708}{{\normalfont
  [arXiv:gr-qc/1909.12708]}}.
\newblock {\url{https://doi.org/10.3847/1538-4357/ac3a6c}}.

\bibitem[Zhou et~al.(2022)Zhou, Zhang, Zhu, and Chang]{Zhou:2021vcw}
Zhou, J.Z.; Zhang, X.; Zhu, Q.H.; Chang, Z.
\newblock {The third order scalar induced gravitational waves}.
\newblock {\em JCAP} {\bf 2022}, {\em 05},~013,
  \href{http://xxx.lanl.gov/abs/2106.01641}{{\normalfont
  [arXiv:astro-ph.CO/2106.01641]}}.
\newblock {\url{https://doi.org/10.1088/1475-7516/2022/05/013}}.

\bibitem[Arun et~al.(2022)]{LISA:2022kgy}
Arun, K.G.;  et~al.
\newblock {New horizons for fundamental physics with LISA}.
\newblock {\em Living Rev. Rel.} {\bf 2022}, {\em 25},~4,
  \href{http://xxx.lanl.gov/abs/2205.01597}{{\normalfont
  [arXiv:gr-qc/2205.01597]}}.
\newblock {\url{https://doi.org/10.1007/s41114-022-00036-9}}.

\bibitem[Zhao and Wang(2022)]{Zhao:2022kvz}
Zhao, Z.C.; Wang, S.
\newblock {Bayesian implications for the primordial black holes from NANOGrav's
  pulsar-timing data by using the scalar induced gravitational waves} {\bf
  2022}.
\newblock  \href{http://xxx.lanl.gov/abs/2211.09450}{{\normalfont
  [arXiv:astro-ph.CO/2211.09450]}}.

\bibitem[Inomata(2021)]{Inomata:2020cck}
Inomata, K.
\newblock {Analytic solutions of scalar perturbations induced by scalar
  perturbations}.
\newblock {\em JCAP} {\bf 2021}, {\em 03},~013,
  \href{http://xxx.lanl.gov/abs/2008.12300}{{\normalfont
  [arXiv:gr-qc/2008.12300]}}.
\newblock {\url{https://doi.org/10.1088/1475-7516/2021/03/013}}.

\bibitem[Carrilho and Malik(2016)]{Carrilho:2015cma}
Carrilho, P.; Malik, K.A.
\newblock {Vector and tensor contributions to the curvature perturbation at
  second order}.
\newblock {\em JCAP} {\bf 2016}, {\em 02},~021,
  \href{http://xxx.lanl.gov/abs/1507.06922}{{\normalfont
  [arXiv:astro-ph.CO/1507.06922]}}.
\newblock {\url{https://doi.org/10.1088/1475-7516/2016/02/021}}.

\bibitem[Zhang et~al.(2017)Zhang, Qin, and Wang]{Zhang:2017hiu}
Zhang, Y.; Qin, F.; Wang, B.
\newblock {Second-order cosmological perturbations. II. Produced by
  scalar-tensor and tensor-tensor couplings}.
\newblock {\em Phys. Rev. D} {\bf 2017}, {\em 96},~103523,
  \href{http://xxx.lanl.gov/abs/1710.06639}{{\normalfont
  [arXiv:gr-qc/1710.06639]}}.
\newblock {\url{https://doi.org/10.1103/PhysRevD.96.103523}}.

\bibitem[Nakama and Suyama(2016)]{Nakama:2016enz}
Nakama, T.; Suyama, T.
\newblock {Primordial black holes as a novel probe of primordial gravitational
  waves. II: Detailed analysis}.
\newblock {\em Phys. Rev. D} {\bf 2016}, {\em 94},~043507,
  \href{http://xxx.lanl.gov/abs/1605.04482}{{\normalfont
  [arXiv:gr-qc/1605.04482]}}.
\newblock {\url{https://doi.org/10.1103/PhysRevD.94.043507}}.

\bibitem[Nakama and Suyama(2015)]{Nakama:2015nea}
Nakama, T.; Suyama, T.
\newblock {Primordial black holes as a novel probe of primordial gravitational
  waves}.
\newblock {\em Phys. Rev. D} {\bf 2015}, {\em 92},~121304,
  \href{http://xxx.lanl.gov/abs/1506.05228}{{\normalfont
  [arXiv:gr-qc/1506.05228]}}.
\newblock {\url{https://doi.org/10.1103/PhysRevD.92.121304}}.

\bibitem[Bari et~al.(2022)Bari, Ricciardone, Bartolo, Bertacca, and
  Matarrese]{Bari:2021xvf}
Bari, P.; Ricciardone, A.; Bartolo, N.; Bertacca, D.; Matarrese, S.
\newblock {Signatures of Primordial Gravitational Waves on the Large-Scale
  Structure of the Universe}.
\newblock {\em Phys. Rev. Lett.} {\bf 2022}, {\em 129},~091301,
  \href{http://xxx.lanl.gov/abs/2111.06884}{{\normalfont
  [arXiv:astro-ph.CO/2111.06884]}}.
\newblock {\url{https://doi.org/10.1103/PhysRevLett.129.091301}}.

\bibitem[Cho et~al.(2020)Cho, Gong, and Oh]{Cho:2020zbh}
Cho, I.; Gong, J.O.; Oh, S.H.
\newblock {Second-order effective energy-momentum tensor of gravitational
  scalar perturbations with perfect fluid}.
\newblock {\em Phys. Rev. D} {\bf 2020}, {\em 102},~043531,
  \href{http://xxx.lanl.gov/abs/2003.12279}{{\normalfont
  [arXiv:gr-qc/2003.12279]}}.
\newblock {\url{https://doi.org/10.1103/PhysRevD.102.043531}}.

\bibitem[Saga(2017)]{Saga:2017hft}
Saga, S.
\newblock {The Vector Mode in the Second-order Cosmological Perturbation
  Theory}.
\newblock PhD thesis, Nagoya U. (main),  2017.
\newblock {\url{https://doi.org/10.1007/978-981-10-8007-4}}.

\bibitem[Lu et~al.(2008)Lu, Ananda, and Clarkson]{Lu:2007cj}
Lu, T.H.C.; Ananda, K.; Clarkson, C.
\newblock {Vector modes generated by primordial density fluctuations}.
\newblock {\em Phys. Rev. D} {\bf 2008}, {\em 77},~043523,
  \href{http://xxx.lanl.gov/abs/0709.1619}{{\normalfont
  [arXiv:astro-ph/0709.1619]}}.
\newblock {\url{https://doi.org/10.1103/PhysRevD.77.043523}}.

\bibitem[Lu et~al.(2009)Lu, Ananda, Clarkson, and Maartens]{Lu:2008ju}
Lu, T.H.C.; Ananda, K.; Clarkson, C.; Maartens, R.
\newblock {The cosmological background of vector modes}.
\newblock {\em JCAP} {\bf 2009}, {\em 02},~023,
  \href{http://xxx.lanl.gov/abs/0812.1349}{{\normalfont
  [arXiv:astro-ph/0812.1349]}}.
\newblock {\url{https://doi.org/10.1088/1475-7516/2009/02/023}}.

\bibitem[Smith et~al.(2008)Smith, Sheth, and Scoccimarro]{Smith:2007sb}
Smith, R.E.; Sheth, R.K.; Scoccimarro, R.
\newblock {An analytic model for the bispectrum of galaxies in redshift space}.
\newblock {\em Phys. Rev. D} {\bf 2008}, {\em 78},~023523,
  \href{http://xxx.lanl.gov/abs/0712.0017}{{\normalfont
  [arXiv:astro-ph/0712.0017]}}.
\newblock {\url{https://doi.org/10.1103/PhysRevD.78.023523}}.

\bibitem[Kamionkowski et~al.(1997)Kamionkowski, Kosowsky, and
  Stebbins]{Kamionkowski:1996zd}
Kamionkowski, M.; Kosowsky, A.; Stebbins, A.
\newblock {A Probe of primordial gravity waves and vorticity}.
\newblock {\em Phys. Rev. Lett.} {\bf 1997}, {\em 78},~2058--2061,
  \href{http://xxx.lanl.gov/abs/astro-ph/9609132}{{\normalfont
  [astro-ph/9609132]}}.
\newblock {\url{https://doi.org/10.1103/PhysRevLett.78.2058}}.

\bibitem[Durrer(1994)]{Durrer:1994uu}
Durrer, R.
\newblock {Light deflection in perturbed Friedmann universes}.
\newblock {\em Phys. Rev. Lett.} {\bf 1994}, {\em 72},~3301--3304,
  \href{http://xxx.lanl.gov/abs/astro-ph/9401033}{{\normalfont
  [astro-ph/9401033]}}.
\newblock {\url{https://doi.org/10.1103/PhysRevLett.72.3301}}.

\bibitem[Yamauchi et~al.(2012)Yamauchi, Namikawa, and Taruya]{Yamauchi:2012bc}
Yamauchi, D.; Namikawa, T.; Taruya, A.
\newblock {Weak lensing generated by vector perturbations and detectability of
  cosmic strings}.
\newblock {\em JCAP} {\bf 2012}, {\em 10},~030,
  \href{http://xxx.lanl.gov/abs/1205.2139}{{\normalfont
  [arXiv:astro-ph.CO/1205.2139]}}.
\newblock {\url{https://doi.org/10.1088/1475-7516/2012/10/030}}.

\bibitem[Saga et~al.(2015)Saga, Yamauchi, and Ichiki]{Saga:2015apa}
Saga, S.; Yamauchi, D.; Ichiki, K.
\newblock {Weak lensing induced by second-order vector mode}.
\newblock {\em Phys. Rev. D} {\bf 2015}, {\em 92},~063533,
  \href{http://xxx.lanl.gov/abs/1505.02774}{{\normalfont
  [arXiv:astro-ph.CO/1505.02774]}}.
\newblock {\url{https://doi.org/10.1103/PhysRevD.92.063533}}.

\bibitem[Chang et~al.(2022)Chang, Zhang, and Zhou]{Chang:2022dhh}
Chang, Z.; Zhang, X.; Zhou, J.Z.
\newblock {The cosmological vector modes from a monochromatic primordial power
  spectrum}.
\newblock {\em JCAP} {\bf 2022}, {\em 10},~084,
  \href{http://xxx.lanl.gov/abs/2207.01231}{{\normalfont
  [arXiv:astro-ph.CO/2207.01231]}}.
\newblock {\url{https://doi.org/10.1088/1475-7516/2022/10/084}}.

\bibitem[Cho et~al.(2022)Cho, Gong, and Oh]{Cho:2022maa}
Cho, I.; Gong, J.O.; Oh, S.H.
\newblock {Second-order energy-momentum tensor of a scalar field}.
\newblock {\em Phys. Rev. D} {\bf 2022}, {\em 106},~084027,
  \href{http://xxx.lanl.gov/abs/2206.11530}{{\normalfont
  [arXiv:gr-qc/2206.11530]}}.
\newblock {\url{https://doi.org/10.1103/PhysRevD.106.084027}}.

\bibitem[Pitrou et~al.(2013)Pitrou, Roy, and Umeh]{Pitrou:2013hga}
Pitrou, C.; Roy, X.; Umeh, O.
\newblock {xPand: An algorithm for perturbing homogeneous cosmologies}.
\newblock {\em Class. Quant. Grav.} {\bf 2013}, {\em 30},~165002,
  \href{http://xxx.lanl.gov/abs/1302.6174}{{\normalfont
  [arXiv:astro-ph.CO/1302.6174]}}.
\newblock {\url{https://doi.org/10.1088/0264-9381/30/16/165002}}.

\bibitem[Bartolo et~al.(2004{\natexlab{a}})Bartolo, Komatsu, Matarrese, and
  Riotto]{Bartolo:2004if}
Bartolo, N.; Komatsu, E.; Matarrese, S.; Riotto, A.
\newblock {Non-Gaussianity from inflation: Theory and observations}.
\newblock {\em Phys. Rept.} {\bf 2004}, {\em 402},~103--266,
  \href{http://xxx.lanl.gov/abs/astro-ph/0406398}{{\normalfont
  [astro-ph/0406398]}}.
\newblock {\url{https://doi.org/10.1016/j.physrep.2004.08.022}}.

\bibitem[Bartolo et~al.(2004{\natexlab{b}})Bartolo, Matarrese, and
  Riotto]{Bartolo:2004ty}
Bartolo, N.; Matarrese, S.; Riotto, A.
\newblock {Gauge-invariant temperature anisotropies and primordial
  non-Gaussianity}.
\newblock {\em Phys. Rev. Lett.} {\bf 2004}, {\em 93},~231301,
  \href{http://xxx.lanl.gov/abs/astro-ph/0407505}{{\normalfont
  [astro-ph/0407505]}}.
\newblock {\url{https://doi.org/10.1103/PhysRevLett.93.231301}}.

\bibitem[Young and Byrnes(2015)]{Young:2015kda}
Young, S.; Byrnes, C.T.
\newblock {Signatures of non-gaussianity in the isocurvature modes of
  primordial black hole dark matter}.
\newblock {\em JCAP} {\bf 2015}, {\em 04},~034,
  \href{http://xxx.lanl.gov/abs/1503.01505}{{\normalfont
  [arXiv:astro-ph.CO/1503.01505]}}.
\newblock {\url{https://doi.org/10.1088/1475-7516/2015/04/034}}.

\end{thebibliography}

\end{document}